\documentclass[prb,twocolumn,showpacs]{revtex4}
\usepackage{graphicx}
\usepackage{amsmath}
\usepackage{amssymb}

\def\lsim{\
  \lower-1.2pt\vbox{\hbox{\rlap{$<$}\lower5pt\vbox{\hbox{$\sim$}}}}\ }
\def\gsim{\
  \lower-1.2pt\vbox{\hbox{\rlap{$>$}\lower5pt\vbox{\hbox{$\sim$}}}}\ }

 \usepackage{amsmath,enumerate, amsfonts, amssymb,amsthm}

\begin{document}
\title{Bose crystal as a standing sound wave}
\author{Maksim Tomchenko} \email{mtomchenko@bitp.kiev.ua}
\affiliation{Bogolyubov Institute for Theoretical Physics, 14-b Metrolohichna Street,
Kiev 03680, Ukraine}

\begin{abstract}
A new class of solutions for Bose crystals with a simple cubic lattice consisting of $N$ atoms is found.
The wave function (WF) of the ground state takes the form
 $\Psi_{0}=e^{S_{w}^{(\textbf{l})}+S_{b}}\prod\limits_{j}\{\sin{(k_{l_{x}}x_{j})}\sin{(k_{l_{y}}y_{j})}
 \sin{(k_{l_{z}}z_{j})}\}$,
 where $e^{S_{b}}$ is the ground-state WF of a fluid,
 $S_{w}^{(\textbf{l})} = \sum\limits_{\textbf{q}\neq 0}S_{1}^{(\textbf{l})}(\textbf{q})\rho_{-\textbf{q}} +
 \sum\limits_{\textbf{q},\textbf{k}\neq 0}^{\textbf{q}+\textbf{k}\neq 0}
 S_{2}^{(\textbf{l})}(\textbf{q},\textbf{k})\rho_{\textbf{k}}\rho_{-\textbf{k}-\textbf{q}}+\ldots $,
  $\rho_{\textbf{k}}=\frac{1}{\sqrt{N}}
   \sum\limits_{j=1}^{N}e^{-i\textbf{k}\textbf{r}_j}$, and $\textbf{k}_{l}=(\pi/a_{l}, \pi/a_{l}, \pi/a_{l})$ ($a_{l}$ is the lattice constant).
   The state with a single longitudinal acoustic phonon is described by the WF
 $\Psi_{\textbf{k}}=\left[\rho_{-\textbf{k}}+
 \sum\limits_{\textbf{q}}P_{2}(\textbf{k},\textbf{q})N^{-1/2}\rho_{\textbf{q}}\rho_{-\textbf{k}-\textbf{q}} +
 \sum\limits_{q_{x}}Q_{1}(q_{x})\rho_{-\textbf{k}-\textbf{q}_{x}} +
 \mbox{7 permutations} \right ]\Psi_{0}$, where the permutations give the terms
 with different signs of components of $\textbf{k}$.
 The structure of $\Psi_{\textbf{k}}$ is such that the excitation corresponds, in fact, to the replacement
of $\textbf{k}_{l}$ in some triple of sines from $\Psi_{0}$ by $\textbf{k}$. Such a structure of $\Psi_{0}$
and $\Psi_{\textbf{k}}$ means that the
crystal is created by sound:
 the ground state of a cubic crystal is formed by $N$ identical three-dimensional standing waves
 similar to a longitudinal sound.
 It is also shown that the crystal in the ground state has a condensate of atoms with $\textbf{k}=\textbf{k}_{l}$.
 The nonclassical inertia moment observed in crystals $He^4$ can be related to the synchronous
 tunneling of condensate atoms.
\end{abstract}

\maketitle

 \section{Introduction}
The science on crystals is developed for many years, and most
properties of crystals are successfully explained. The current
interest is focused on the regions of unconventional and yet
unguessed properties manifesting themselves, in particular, at
extra-low temperatures (see surveys \cite{rev1,rev2,rev3}). For
the crystals with a charged lattice and the Bose crystals, such
regions are, respectively, high-temperature superconductivity and
supersolid phenomena. The splash-up of interest arose after the
excellent experiments by E. Kim and M. Chan
\cite{kim-chan1,kim-chan2}, where a nonclassical inertia moment
(NCIM) of a crystal $He^4$ was discovered. Later on, a number of
new interesting properties joined by the term ``supersolid'' were
found
\cite{rittner2006,beamish2006,sasaki2007,day2007,chan2007,ftint2008,reppy2008,sf-sp1,sf-sp2,chan2009,kim2011}.
There are almost no doubts that the ``supersolid'' phenomenon is
related to the superfluidity of quantum crystals, which was
predicted long ago \cite{andreev1969}. However, the physical
nature of the superfluidity and the ``supersolid'' phenomenon is
not clear yet, though a lot of models were developed
\cite{shevch1987,ceperley2004,spigs2006,superglass,grain,disloc-glass,disloc-net,krai2011,rota2011,andreev2011,anderson2011}.

The results obtained below indicate that the basic property of crystals, i.e.,
the nature of crystalline ordering, is not completely clear as well.

Commonly accepted is the following structure of the WF of a Bose
crystal \cite{saunders1962,mullin1964,nosanow1964,nosanow1966}:
   \begin{equation}
   \Psi_{0} = e^{-\sum\limits_{i>j}S_{J}(\textbf{r}_{i}-\textbf{r}_{j})}\prod\limits_{i=1}^{N} \varphi(\textbf{r}_{i}-\textbf{R}_{i}),
 \label{1-1}    \end{equation}
where $N$ is the total number of atoms of a crystal,
$\textbf{r}_j$ and $\textbf{R}_j$ are coordinates of atoms and
sites of the lattice, the exponential function is the Bijl-Jastrow
function taking correlations into account, and
$\varphi(\textbf{r})$ is usually written in the approximation of
small oscillations: $\varphi(\textbf{r})=e^{-\alpha^{2} r^{2}/2}$.
  By the modern ideas, a crystal is formed since its energy is less than that of a fluid.

In what follows, we propose the basically new wave solution for
the WF of a Bose crystal. In Ref.~\onlinecite{zero-liquid} (cited below
as I), it was shown that the states of a system of $N$ interacting
Bose particles positioned in a rectangular box $L_{x}\times
L_{y}\times L_{z}$ in size include the states with the WF
 \begin{equation}
 \Psi_{0}=Ae^{S_{w}^{(\textbf{l})}+\tilde{S}_{b}}\prod\limits_{j=1}^{N}\{\sin{(k_{l_{x}}x_{j})}\sin{(k_{l_{y}}y_{j})}
 \sin{(k_{l_{z}}z_{j})}\},
  \label{1-6}    \end{equation}
where $(k_{l_{x}}, k_{l_{y}}, k_{l_{z}})\equiv
\textbf{k}_{\textbf{l}}=(l_{x}\pi/L_{x}, l_{y}\pi/L_{y},
l_{z}\pi/L_{z})$, $l_{x}, l_{y}, l_{z}$ are integers, and the
remaining designations are given in Sec. 2.
 In I, we analyzed $\Psi_{0}$ only with  $\textbf{k}_{\textbf{l}}=\textbf{k}_{1}=(\pi/L_{x}, \pi/L_{y}, \pi/L_{z})$
 (here, $l_{x}, l_{y}, l_{z}=1$),
which describes the ground state of a gas and a fluid. In this
case, the sines in (\ref{1-6}) form a standing half-wave that
covers the whole system and rests on the boundaries. But, $l_{x}, l_{y}$ and $l_{z}$
may obviously take in the solution any other integer values except zero.  
It is natural to assume that
 if the half-wave is equal to the lattice constant, then WF (\ref{1-6})
 describes a rectangular crystal lattice.
Moreover, (\ref{1-6}) is one of the exact solutions of the
Schr\"{o}dinger equation with zero boundary conditions (BCs).
 In what follows, we will study this solution and the solution for a longitudinal acoustic phonon. We will show that the solutions
 agree with observable properties of crystals and predict a number of specific features, in particular,
the condensates of phonons and atoms in a crystal. Solution
(\ref{1-6}) testifies to the \textit{wave nature of a crystal}. I
did not deal  with crystals earlier and have found the solutions
accidentally, while studying the
 microstructure of a fluid.

 A short announcement of the results will be published separately \cite{nature1}.

 \section{Ground state of a crystal}
 Due to the presence of a product of sines in $\Psi_{0}$ (\ref{1-6}),
the system can be partitioned into $l_{x}\times l_{y}\times l_{z}$
identical domains separated by plane surfaces, on which
$\Psi_{0}=0$. We may assume that if each domain contains one atom,
and the size of domains is close to the equilibrium interatomic
distance, then the system is stable. A crystal corresponds,
obviously, to a system of domains with the size at which the
energy of the system is minimum. If the number of domains is equal
to the number of atoms, then $l_{x}=L_{x}/\bar{R}_{x}$,
$l_{y}=L_{y}/\bar{R}_{y}$, $l_{z}=L_{z}/\bar{R}_{z}$, where
$\bar{R}_{x}$, $\bar{R}_{y}$ and $\bar{R}_{z}$ are the periods of
the lattice along the appropriate axes. For a cubic crystal, they
are equal to $\bar{R}$. Let us study the properties of such system
and make clear whether they correspond to the properties of
crystals. We consider only a simple cubic (sc) lattice.

Consider WF (\ref{1-6}). In it, $\tilde{S}_{b}$ is given by the formula (see I)
\begin{eqnarray}
& \tilde{S}_{b}& =
 \sum\limits_{\textbf{k}_{1}\neq 0}^{(\pi)}\frac{a_{2}(\textbf{k}_{1})}{2!}\rho_{\textbf{k}_{1}}
   \rho_{-\textbf{k}_{1}} \nonumber\\  &+&
 \sum\limits_{\textbf{k}_{1},\textbf{k}_{2}\neq 0}^{(\pi)\,\textbf{k}_{1}+\textbf{k}_{2}\not= 0}
  \frac{a_{3}(\textbf{k}_{1},\textbf{k}_{2})}{3!\sqrt{N}}
 \rho_{\textbf{k}_{1}}\rho_{\textbf{k}_{2}}\rho_{-\textbf{k}_{1} - \textbf{k}_{2}}+\ldots,
   \label{2-1} \end{eqnarray}
where
\begin{equation}
   \rho_{\textbf{k}} = \frac{1}{\sqrt{N}}
   \sum\limits_{j=1}^{N}e^{-i\textbf{k}\textbf{r}_j} \quad (\textbf{k}\not= 0),
 \label{1-4}    \end{equation}
and the summation is carried on over the wave vectors
\begin{equation}
  \textbf{k}=\pi \left (\frac{j_{x}}{L_{x}}, \frac{j_{y}}{L_{y}},
 \frac{j_{z}}{L_{z}} \right )
    \label{2-2} \end{equation}
($j_{x}, j_{y}, j_{z}$ are integers), which are multiple to
$\pi/L.$ This is denoted by the symbol $(\pi)$ above the sums
(under cyclic BCs, $\textbf{k}$ are multiple to
$2\pi/L,$ and the solutions are strongly changed). The function
$S_{w}^{(\textbf{l})}$ in (\ref{1-6}) takes the form of an
infinite series (see I)
\begin{eqnarray}
& S_{w}^{(\textbf{l})} &= \sum\limits_{\textbf{q}\neq 0}S_{1}^{(\textbf{l})}(\textbf{q})\rho_{-\textbf{q}} +
 \sum\limits_{\textbf{q},\textbf{k}_{1}\neq 0}^{\textbf{q}+\textbf{k}_{1}\neq 0}
 \frac{S_{2}^{(\textbf{l})}(\textbf{q},\textbf{k}_{1})}{\sqrt{N}}\rho_{\textbf{k}_{1}}\rho_{-\textbf{k}_{1}-\textbf{q}}+\nonumber \\
 &+& \sum\limits_{\textbf{q},\textbf{k}_{1},\textbf{k}_{2}\neq 0}^{\textbf{q}+
 \textbf{k}_{1}+\textbf{k}_{2}\neq 0}
 \frac{S_{3}^{(\textbf{l})}(\textbf{q},\textbf{k}_{1},\textbf{k}_{2})}{N}
 \rho_{\textbf{k}_{1}}\rho_{\textbf{k}_{2}}
 \rho_{-\textbf{k}_{1}-\textbf{k}_{2}-\textbf{q}}+\ldots
     \label{2-3} \end{eqnarray}
where $\textbf{k}_{j}$ run values (\ref{2-2}), and $\textbf{q}$
take values $2\pi \left (\frac{j_{x}}{L_{x}}, \frac{j_{y}}{L_{y}},
 \frac{j_{z}}{L_{z}} \right )$. In (\ref{2-1}),
 the first sum is the Bijl-Jastrow function written in the variables $\rho_{\textbf{k}}$.

Let the faces of the crystal be ideally plane and parallel to atomic planes of the lattice.
Every face creates a potential barrier for atoms of the crystal.
We model this barrier for the face that is perpendicular to the $X$ and has the coordinate $x=0$ by a step
 \begin{equation}
 U_{w}(x) \approx
\left [ \begin{array}{ccc}
    U_{s}  & \   x\leq 0,   & \\
    0  & \ x>0. &
\label{2-4} \end{array} \right. \end{equation} The potential of
the face with the coordinate $x=L_{x}$ is $U_{w}(L_{x}-x)$.
Analogously, we can consider four other faces. For simplicity, we
take $U_{w}=\infty$, which corresponds to zero BCs.

In I,  the product of assigning sines
\begin{equation}
\Psi^{bare}_{sc}(N)=\prod\limits_{j=1}^{N}\{\sin{(k_{l_{x}}x_{j})}
\sin{(k_{l_{y}}y_{j})}\sin{(k_{l_{z}}z_{j})}\}
     \label{2-5} \end{equation}
has been factor out from the equations for $\Psi_{0}$ and
$\Psi_{\textbf{k}}$. In this case, we obtain the cotangent
$\cot{(\pi l_{x} x/L_{x})}$, which cannot be expanded in a Fourier
series, because $\int\limits_{0}^{L_{x}}|\cot{(\pi l_{x}
x/L_{x})}|dx=\infty$. To overcome this problem, we will use the
following. It is easy to see that, under the change $\sin{kx}
\rightarrow |\sin{kx}|,$ the derivative
$(d|\sin{kx}|/dx)(1/|\sin{kx}|)$ gives again $\cot{(kx)}$.
Therefore, we replace $\Psi^{bare}_{sc}$ (\ref{2-5}) by the WF
\begin{equation}
\Psi^{bare}_{sc}(N)=\prod\limits_{j=1}^{N}\{|\sin{(k_{l_{x}}x_{j})}|\times
|\sin{(k_{l_{y}}y_{j})}|\times|\sin{(k_{l_{z}}z_{j})}|\}
     \label{2-6} \end{equation}
and pass to
\begin{eqnarray}
\Psi^{bare}_{sc}(N)&=&\prod\limits_{j=1}^{N}\left \{\left (|\sin{(k_{l_{x}}x_{j})}|+\delta \right )
\right. \label{2-7}  \\ &\times& \left.
\left (|\sin{(k_{l_{y}}y_{j})}|+\delta \right )\times \left (|\sin{(k_{l_{z}}z_{j})}|+\delta \right )\right \},
   \nonumber   \end{eqnarray}
where $ \delta > 0, \delta \rightarrow 0$. For such $\Psi^{bare}_{sc},$ we obtain the function
\begin{equation}
\tilde{f}(x)=(df/dx)(1/f), \quad f=|\sin{(k_{l_{x}}x_{j})}|+\delta
     \label{2-f} \end{equation}
instead of $\cot{(k_{l_{x}}x)}.$ The former can be expanded in a
Fourier series. For the singular point $x_{0},$ the series gives
the arithmetic mean value of those at the points $x_{0}-0$ and
$x_{0}+0$. In the final formulas, we transit to the limit $\delta
\rightarrow 0$, which returns us to functions (\ref{2-6}) and
(\ref{2-5}). Bearing this fact in mind, we will immediately use in
formulas the expansions at $\delta = 0$:
\begin{equation}
 \cot{(k_{l_{x}}x)} \equiv \cot{(\pi l_{x} x/L_{x})} =
 \sum\limits_{j_{x}}C_{l_{x}}(q_{x})e^{iq_{x}x}.
     \label{2-8} \end{equation}
Here, $q_{x}=2\pi j_{x}/L_{x} $, $j_{x}$ runs all integers, and
 \begin{equation}
  C_{l_{x}}(q_{x}) \equiv C_{l_{x}}(j_{x}) =
\left [ \begin{array}{ccc}
    -i   & \   \mbox{for}\ j_{x}= l_{x},  2l_{x},  3l_{x}, \ldots  & \\
     i   & \   \mbox{for}\ j_{x}=-l_{x}, -2l_{x}, -3l_{x}, \ldots  & \\
   0    & \ \mbox{for the rest}\ j_{x}. &
   \label{2-9} \end{array} \right.  \end{equation}
The proof of formulas (\ref{2-8}) and (\ref{2-9}) is given in Appendix.

We now have all the required  in order to write the WFs of a Bose
crystal.  The WF of the ground state is set by formula
(\ref{1-6}), where $\tilde{S}_{b}$ and $S_{w}^{(\textbf{l})}$ are
given by formulas (\ref{2-1}) and (\ref{2-3}). The equations for
the functions $a_{l}$ and the ground-state energy of an atom
($E_{0}$) follow from equations in I with the changes
$\textbf{k}_{1}\rightarrow \textbf{k}_{\textbf{l}}\equiv
\textbf{k}_{c}$, $C_{1}(q)\rightarrow C_{l}(q)$, $S_{j}^{(1)}
\rightarrow S_{j}^{(\textbf{l})}$:
\begin{equation}
 E_{0}= \tilde{E}^{b}_0 + A_{1},
     \label{2-10} \end{equation}
 \begin{eqnarray}
& A_{1}&=  \frac{\hbar^2}{2m}\left [ k_{c}^{2} -
 \frac{1}{N}\sum\limits_{\textbf{q}\neq 0}^{(2\pi)}q^{2}S_{1}^{(\textbf{l})}(\textbf{q})S_{1}^{(\textbf{l})}(-\textbf{q})
  \right. \label{2-11} \\
 &-& \left. \frac{i}{\sqrt{N}}\sum\limits_{q_{x}\neq 0}^{(2\pi)}2k_{cx}q_{x}C_{l_{x}}(q_{x})S_{1}^{(\textbf{l})}(-\textbf{q}_{x})
 + (x\rightarrow y, z)\right ],
     \nonumber \end{eqnarray}
 \begin{equation}
 \tilde{E}^{b}_0 = \frac{N-1}{2^{d+1}N}n\nu_{3}(0) - \frac{1}{2^{d+1}N}
 \sum\limits_{\textbf{k}\not= 0}^{(\pi)}n\nu_{3}(k) - \frac{1}{N }
 \sum\limits_{\textbf{k}\not= 0}^{(\pi)}\frac{\hbar^{2}k^2}{2m}a_{2}(\textbf{k}),
     \label{2-12} \end{equation}
 \begin{eqnarray}
&&\frac{n\nu_{3}(k)m}{2^{d}\hbar^2} + a_{2}(\textbf{k})k^2   -
 a_{2}^{2}(\textbf{k})k^2 = A_{2}(\textbf{k}) \label{2-13} \\ &+&
 \frac{1}{N}\sum\limits_{\textbf{q}\not= 0}^{(\pi)}a_{3}(\textbf{k},\textbf{q})
 (q^2+\textbf{k}\textbf{q}) + \frac{1}{2N}
  \sum\limits_{\textbf{q}\not= 0}^{(\pi)} q^2
a_{4}(\textbf{q},-\textbf{q},\textbf{k}),
  \nonumber \end{eqnarray}
 \begin{equation}
 a_{3}(\textbf{k}, \textbf{q}) \approx A_{3}(\textbf{k},\textbf{q}) -2\frac{R(\textbf{k},\textbf{q})}{\epsilon_{0}(k)+\epsilon_{0}(q)+
 \epsilon_{0}(\textbf{k}+\textbf{q})},
     \label{2-14} \end{equation}
     \begin{eqnarray}
 R(\textbf{k}, \textbf{q}) &=& \textbf{k}\textbf{q}a_{2}(\textbf{k})
 a_{2}(\textbf{q})-\textbf{k}(\textbf{k}+\textbf{q})a_{2}(\textbf{k})
 a_{2}(\textbf{k}+\textbf{q}) \nonumber \\ &-&\textbf{q}(\textbf{k}+\textbf{q})a_{2}(\textbf{q})
 a_{2}(\textbf{k}+\textbf{q}),
     \label{1-a3r} \end{eqnarray}
     \begin{eqnarray}
A_{2}(\textbf{k}) &=& \frac{1}{\sqrt{N}}\sum\limits_{q_{x}\neq 0}^{(2\pi)}ik_{cx}C_{l_{x}}(-q_{x})\left
[4(q_{x}+k_{x})S_{2}^{(\textbf{l})}(\textbf{q}_{x},\textbf{k})\right. \nonumber\\
&+& \left. 6q_{x}S_{3}^{(\textbf{l})}(\textbf{q}_{x},-\textbf{q}_{x},\textbf{k})
\right ] +(x \rightarrow y, z)  \nonumber\\
&+&  \frac{1}{N}\sum\limits_{\textbf{q}\neq 0}^{(2\pi)}\left [4(q^{2}+\textbf{q}\textbf{k})S_{1}^{(\textbf{l})}(-\textbf{q})
S_{2}^{(\textbf{l})}(\textbf{q},\textbf{k})+ \right. \nonumber\\ &+&
 4(\textbf{k}+\textbf{q})^{2}S_{2}^{(\textbf{l})}(\textbf{q},\textbf{k})S_{2}^{(\textbf{l})}(-\textbf{q},-\textbf{k}) \nonumber\\
 &+& \left. 6q^{2}S_{1}^{(\textbf{l})}(-\textbf{q}) S_{3}^{(\textbf{l})}(\textbf{q},\textbf{k},-\textbf{k})\right ],
 \label{2-15}  \end{eqnarray}
 where $\epsilon_{0}(\textbf{k})=k^{2}(1-2a_{2}(\textbf{k}))$,
 $d=3$ is the dimension of the system, and
 \begin{equation}
  \nu_{3}(k) = \int U_{3}(r)e^{-i\textbf{k}\textbf{r}}d\textbf{r}
    \label{2-15b} \end{equation}
is the Fourier transform of the interaction potential $U_{3}(r)$ of two Bose particles.
The equations for the functions $S_{j}^{(\textbf{l})}$
can be obtained analogously from the equations for $S_{j}^{(1)}$ (see I):
  \begin{eqnarray}
&& S_{1}^{(\textbf{l})}(\textbf{q})\epsilon_{0}(\textbf{q})= \nonumber\\ &=& -i\sqrt{N} 2k_{cx}q_{x}a_{2}(-\textbf{q}_{x})C_{l_{x}}(q_{x})\delta_{\textbf{q},\textbf{q}_{x}}
 + (x\rightarrow y, z)   \nonumber\\ &+& \frac{1}{N}\sum\limits_{\textbf{q}_{1}\neq 0}^{(\pi)}\left \{2(q_{1}^{2}+\textbf{q}_{1}\textbf{q})
 S_{2}^{(\textbf{l})}(\textbf{q},\textbf{q}_{1})+
6q_{1}^{2}S_{3}^{(\textbf{l})}(\textbf{q},\textbf{q}_{1},-\textbf{q}_{1}) \right. \nonumber\\ &+&
\sqrt{N}4q_{1}^{2}S_{1}^{(\textbf{l})}(\textbf{q}_{1})S_{2}^{(\textbf{l})}(\textbf{q}-\textbf{q}_{1},\textbf{q}_{1})
\nonumber\\
 &+&\left. \sqrt{N}(q_{1}^{2}-\textbf{q}_{1}\textbf{q})S_{1}^{(\textbf{l})}(\textbf{q}_{1})S_{1}^{(\textbf{l})}(\textbf{q}-\textbf{q}_{1}) \right \}\nonumber\\ &+&
 \left \{ \sum\limits_{p_{x}\neq 0}^{(2\pi)}2ik_{cx}C_{l_{x}}(p_{x})
 \left [(q_{x}-p_{x})S_{1}^{(\textbf{l})}(\textbf{q}-\textbf{p}_{x}) \right. \right. \nonumber\\
 & -& \left.\left. 2p_{x}S_{2}^{(\textbf{l})}(\textbf{q}-\textbf{p}_{x},-\textbf{q})
 \right ] + (x\rightarrow y, z) \right \},
 \label{2-16}  \end{eqnarray}
\begin{eqnarray}
&& S_{2}^{(\textbf{l})}(\textbf{q},\textbf{q}_{1})\left [\epsilon_{0}(\textbf{q}_{1}) + \epsilon_{0}(\textbf{q}+\textbf{q}_{1})
\right ]\nonumber\\ &+&
2S_{1}^{(\textbf{l})}(\textbf{q})a_{2}(-\textbf{q}_{1})\textbf{q}\textbf{q}_{1}-q^{2}S_{1}^{(\textbf{l})}(\textbf{q})a_{3}(\textbf{q},\textbf{q}_{1})
 \nonumber\\ &=& \delta_{\textbf{q},\textbf{q}_{x}}\sqrt{N}ik_{cx}C_{l_{x}}(q_{x})\left \{ 2q_{1x}a_{2}(\textbf{q}_{1})-
q_{x}a_{3}(\textbf{q}_{1},\textbf{q}_{x})\right \}
 \nonumber\\ &&
 +\sum\limits_{p_{x}\neq 0}^{(2\pi)}ik_{cx}C_{l_{x}}(p_{x})\left [4(q_{x}+q_{1x}-p_{x})
 S_{2}^{(\textbf{l})}(\textbf{q}-\textbf{p}_{x},\textbf{q}_{1})\right. \nonumber\\ &-& \left. 6p_{x}S_{3}^{(\textbf{l})}(\textbf{q}-\textbf{p}_{x},\textbf{q}_{1},-\textbf{q}-\textbf{q}_{1})
\right ] + (x\rightarrow y, z) \nonumber\\ &+& \frac{1}{\sqrt{N}}\sum\limits_{\textbf{q}_{2}\neq 0}^{(\pi)}\left \{
\frac{2}{\sqrt{N}}(q_{2}^{2}-\textbf{q}_{1}\textbf{q}_{2})S_{3}^{(\textbf{l})}(\textbf{q},\textbf{q}_{1}-\textbf{q}_{2},\textbf{q}_{2}) \right.
\nonumber\\ &+&
\frac{4}{\sqrt{N}}\textbf{q}_{2}(\textbf{q}_{1}+\textbf{q}_{2}+\textbf{q})S_{3}^{(\textbf{l})}(\textbf{q},\textbf{q}_{1},\textbf{q}_{2})
\nonumber\\ &-& 4\textbf{q}_{2}(\textbf{q}_{1}-\textbf{q}_{2}+\textbf{q})S_{1}^{(\textbf{l})}(\textbf{q}_{2})S_{2}^{(\textbf{l})}(\textbf{q}-\textbf{q}_{2},\textbf{q}_{1})
\nonumber\\ &+& 6q_{2}^{2}S_{1}^{(\textbf{l})}(\textbf{q}_{2})S_{3}^{(\textbf{l})}(\textbf{q}-\textbf{q}_{2},\textbf{q}_{1},-\textbf{q}-\textbf{q}_{1})
\nonumber\\ &+& \left. 4(\textbf{q}_{1}+\textbf{q}_{2})^{2}S_{2}^{(\textbf{l})}(\textbf{q}_{2},\textbf{q}_{1})S_{2}^{(\textbf{l})}(\textbf{q}-\textbf{q}_{2},\textbf{q}_{1}+\textbf{q}_{2})
\right \}.
 \label{2-17}  \end{eqnarray}

Equations (\ref{2-10})--(\ref{2-17}) are a complicated system of
nonlinear integral equations, whose solutions determine the
properties of the ground state of the crystal with a rectangular
lattice. Let us analyze these equations. Of a paramount interest
are the value of $E_{0}$ and the distribution of atoms in the
crystal.

It is seen from Eq. (\ref{2-16}) that the nonzero value of $S_{1}^{(\textbf{l})}(\textbf{q})$
is determined by the first ``one-dimensional'' term  $\sim C_{l_{x}}(q_{x})\delta_{\textbf{q},\textbf{q}_{x}}$ on the right-hand side.
This yields the one-dimensional
solutions $S_{1}^{(\textbf{l})}(\textbf{q}_{x})\delta_{\textbf{q},\textbf{q}_{x}}$,
$S_{1}^{(\textbf{l})}(\textbf{q}_{y})\delta_{\textbf{q},\textbf{q}_{y}},$ and
$S_{1}^{(\textbf{l})}(\textbf{q}_{z})\delta_{\textbf{q},\textbf{q}_{z}}$.
However, the terms $\sim S_{1}^{(\textbf{l})}(\textbf{q}_{1})S_{2}^{(1)}(\textbf{q}-\textbf{q}_{1},\textbf{q}_{1})$
and some other ones on the right-hand side generate also not one-dimensional solutions of the form
$S_{1}^{(\textbf{l})}(\textbf{q}_{x}+\textbf{q}_{y})\delta_{\textbf{q},\textbf{q}_{x}+\textbf{q}_{y}}$
and $S_{1}^{(\textbf{l})}(\textbf{q}_{x}+\textbf{q}_{y}+\textbf{q}_{z})$ of the same order ($\sim \sqrt{N}$)
as one-dimensional solutions.  Since  $C_{l_{x}}(j_{x})$ are nonzero only at the ``resonance'' points
$j_{x}= \pm l_{x}, \pm 2l_{x}, \pm 3l_{x}, \ldots $,
the function  $S_{1}^{(\textbf{l})}(\textbf{q})$ is nonzero only for the ``resonance''  wave vectors
 \begin{equation}
  \textbf{q}^{res}(\textbf{r})=2\pi \left (\frac{r_{x} l_{x}}{L_{x}}, \frac{r_{y} l_{y}}{L_{y}},
  \frac{r_{z} l_{z}}{L_{z}} \right ),
    \label{2-19} \end{equation}
where   $\textbf{r}=(r_{x}, r_{y}, r_{z})$, $r_{x}, r_{y}, r_{z} = \pm 1, \pm 2, \pm 3, \ldots$.
With regard for these relations, we can write the solution
 \begin{eqnarray}
&&S_{1}^{(\textbf{l})}(\textbf{q}=\textbf{q}_{x}+\textbf{q}_{y}+\textbf{q}_{z})=
S_{1}^{(\textbf{l})}(\textbf{q})\left \{\delta_{\textbf{q},\textbf{q}^{res}_{x}}+
\delta_{\textbf{q},\textbf{q}^{res}_{y}} \right. \nonumber \\ &+& 
\delta_{\textbf{q},\textbf{q}^{res}_{z}} +
\delta_{\textbf{q},\textbf{q}^{res}_{x}+\textbf{q}^{res}_{y}}  +
\delta_{\textbf{q},\textbf{q}^{res}_{x}+\textbf{q}^{res}_{z}} +
\delta_{\textbf{q},\textbf{q}^{res}_{z}+\textbf{q}^{res}_{y}} \nonumber \\ &+&
\left. \delta_{\textbf{q},\textbf{q}^{res}}\right \}
      \label{2-20} \end{eqnarray}
and, analogously, $S_{j\geq 2}^{(\textbf{l})}(\textbf{q},\textbf{q}_{1},...\textbf{q}_{j-1}).$
As is seen from Eq. (\ref{2-16}), the values of $S_{1}^{(\textbf{l})}(\textbf{q}_{x})$ depend significantly on the
sums with $S_{2}^{(\textbf{l})}$, $S_{3}^{(\textbf{l})}$
and with the terms $\sim S_{1}^{(\textbf{l})}S_{1}^{(\textbf{l})},$ and $\sim S_{1}^{(\textbf{l})}S_{2}^{(\textbf{l})}$.
However, we restrict ourselves to the
zero approximation
\begin{equation}
 S_{1}^{(\textbf{l})}(\textbf{q}_{x}) \approx -\frac{i\sqrt{N}2k_{cx}C_{l_{x}}(q_{x})a_{2}(-\textbf{q}_{x})\delta_{\textbf{q}_{x},\textbf{q}^{res}_{x}}}{q_{x}-2q_{x}a_{2}(\textbf{q}_{x})},
      \label{2-21} \end{equation}
 \begin{eqnarray}
 S_{2}^{(\textbf{l})}(\textbf{q}_{x},\textbf{q}_{1})& \approx & \delta_{\textbf{q}_{x},\textbf{q}^{res}_{x}}
 \frac{i\sqrt{N}k_{cx}C_{l_{x}}(q_{x})}{1-2a_{2}(\textbf{q}_{x})} \nonumber \\
 &\times & \frac{2q_{1x}a_{2}(\textbf{q}_{1})-q_{x}a_{3}(\textbf{q}_{1},\textbf{q}_{x})}{
 \epsilon_{0}(\textbf{q}_{1}) + \epsilon_{0}(\textbf{q}_{x}+\textbf{q}_{1})}.
      \label{2-22} \end{eqnarray}
The analogous relations are true for $S_{1}^{(\textbf{l})}(\textbf{q}_{y})$, $S_{1}^{(\textbf{l})}(\textbf{q}_{z})$ and
$S_{2}^{(\textbf{l})}(\textbf{q}_{y},\textbf{q}_{1})$, $S_{2}^{(\textbf{l})}(\textbf{q}_{z},\textbf{q}_{1})$.
The omitted corrections can renormalize $S_{1}^{(\textbf{l})}(\textbf{q}_{x}),$ by changing its value by several times.
In this case, the higher corrections are damped by the decrease of $a_{2}(\textbf{q})$ with increase in $q$.
For the not one-dimensional solutions  $S_{1}^{(\textbf{l})}(\textbf{q}_{x}+\textbf{q}_{y})$ and
$S_{1}^{(\textbf{l})}(\textbf{q}_{x}+\textbf{q}_{y}+\textbf{q}_{z}),$ even the zero approximation is a complicated
sum (over $\textbf{q}_{1}$) of the terms
$\sim S_{1}^{(\textbf{l})}(\textbf{q}_{1})S_{1}^{(\textbf{l})}(\textbf{q}-\textbf{q}_{1})$
and $\sim  S_{1}^{(\textbf{l})}(\textbf{q}_{1})S_{2}^{(\textbf{l})}(\textbf{q}-\textbf{q}_{1},\textbf{q}_{1})$.
By our estimates, the not one-dimensional solutions are significantly less than the one-dimensional ones.
Due to the complexity of the equations, we neglect the not one-dimensional solutions:
\begin{equation}
S_{1}^{(\textbf{l})}(\textbf{q}) \simeq
S_{1}^{(\textbf{l})}(\textbf{q})\left \{\delta_{\textbf{q},\textbf{q}^{res}_{x}}+
\delta_{\textbf{q},\textbf{q}^{res}_{y}} +
\delta_{\textbf{q},\textbf{q}^{res}_{z}}  \right \}.
      \label{2-23} \end{equation}

Let us study the distribution of atoms along the $x$-axis
(coinciding with one of the axes of a crystal) for the lattice of
$He^4$ atoms at
$\bar{R}_{x}=\bar{R}_{y}=\bar{R}_{z}=\bar{R}=3.6\,\mbox{\AA}$.
According to I, the probability for an atom to be at a point $x$
is approximately determined by the formulas
\begin{equation}
 \psi(x) \simeq \sin{(k_{cx}x)}e^{s_{1}(x)},
      \label{2-24} \end{equation}
\begin{equation}
 s_{1}(x) \simeq \sum\limits_{q_{x}=2\pi j_{x}/L_{x}}^{(2\pi)}\frac{S_{1}^{(\textbf{l})}(\textbf{q}_{x})}{\sqrt{N}}e^{iq_{x}x},
      \label{2-25} \end{equation}
where $j_{x}=\pm 1, \pm 2, \pm 3,  \ldots$ and $k_{cx} =\pi /\bar{R}$.
Despite the approximate character of the formulas, we may expect that they give the general form of a probability
distribution in a cell.

Since $S_{1}^{(\textbf{l})}(\textbf{q}_{x})\neq 0$ only at the resonance points and
$S_{1}^{(\textbf{l})}(-\textbf{q}_{x})=S_{1}^{(\textbf{l})}(\textbf{q}_{x})$, we have
\begin{equation}
 s_{1}(x) \simeq \sum\limits_{r_{x}= 1, 2, \ldots}\frac{S_{1}^{(\textbf{l})}(\textbf{q}_{x})}{\sqrt{N}}2\cos{(q_{x}x)},
      \label{2-26} \end{equation}
where $q_{x}=r_{x} 2\pi l_{x}/L_{x}=r_{x} 2\pi /\bar{R}$. We will determine $S_{1}^{(\textbf{l})}(\textbf{q}_{x})$ from Eq. (\ref{2-21}), by using the
zero approximation for $a_{2}(q),$
\begin{equation}
 a_{2}(q) = \frac{1}{2}- \sqrt{\frac{1}{4}+\frac{n \nu_{3}(q)m}{8\hbar^{2}q^{2}}},
      \label{2-27} \end{equation}
which follows from (\ref{2-13}) if all sums are neglected. Note
that we took the solution with the sign ``minus'' before the root
(see I).

We choose the interatomic interaction potential for atoms of the
crystal, as in I:
 \begin{equation}
 U_{3}(\textbf{r}) \approx
\left [ \begin{array}{ccc}
    U_{b}  & \   r\leq a  & \\
    U_{bd} & \   a\leq r \leq b  & \\
    0  & \ r>b. &
\label{2-28} \end{array} \right. \end{equation} Below, we use,
unless otherwise indicated,   $a= 2\,\mbox{\AA}$, $b=
4\,\mbox{\AA}$,  $U_{bd}=-9\,$K,  which corresponds approximately
to $He^4$ atoms. This potential is sufficiently crude, but it is
qualitatively proper and has the analytical Fourier transform
\begin{equation}
 \nu_{3}(k)  = \frac{4\pi }{k^3}\tilde{\nu}_{3}(k),
       \label{2-29} \end{equation}
 \begin{eqnarray}
 \tilde{\nu}_{3}(k) & =& [U_{b}-U_{bd}][\sin{(ak)}-ak\cos{(ak)]}  \nonumber \\
 &+&  U_{bd}[\sin{(bk)}-bk\cos{(bk)}].
      \label{2-30} \end{eqnarray}

      \begin{figure}[h]
\centerline{\includegraphics[width=85mm]{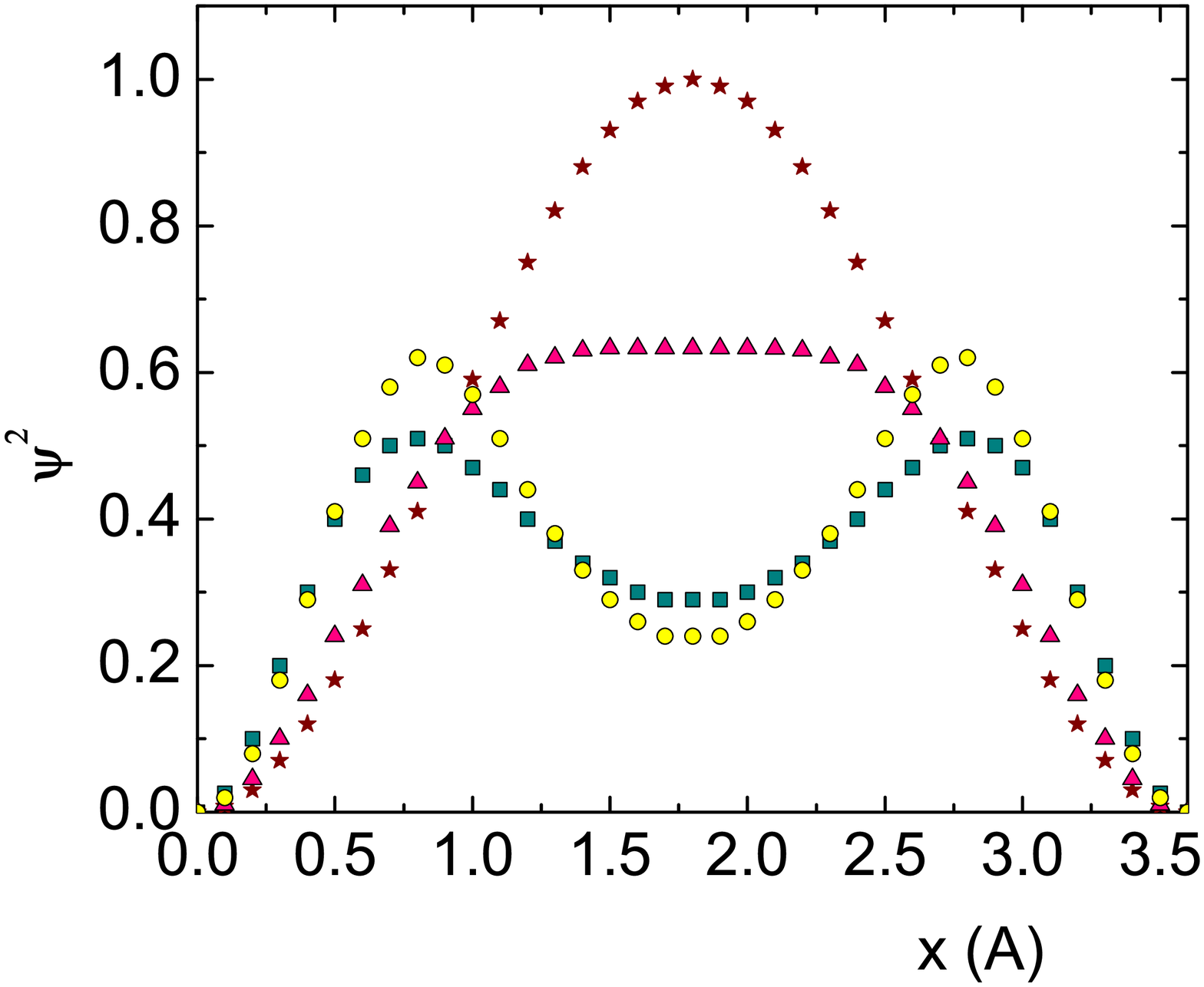}}
\caption{Probability density $|\psi(x)|^{2}$  (\ref{2-24}), (\ref{2-26}) versus the coordinate $x$ of a cell
of the sc crystal at the potential with $a=2\,\mbox{\AA}$, $b=4\,\mbox{\AA}$,  $U_{bd} =-9\,$K
and the barriers $U_{b} =300\,$K (triangles), $U_{b} =3000\,$K (circles),
and $U_{b} =3000\,$K with regard for only the first term in sum (\ref{2-26}) (squares).
 Stars stand for the bare function $\sin^{2}{(k_{cx}x)}$. The $x$-axis  coincides with
 the $x$-axis of a cell of the crystal,
 values of $x$ are given in {\AA}. \label{fig1}}
\end{figure}
By formulas (\ref{2-24})--(\ref{2-30}), we determine $\psi(x)$
(\ref{2-24}). It has a periodic shape, in correspondence with the
domain structure. The distribution $|\psi(x)|^2$ in one of the
domains is shown in Fig.~\ref{fig1}. The value of $\bar{R}$ is
chosen like that for He II: $\bar{R} = 3.6\,\mbox{\AA}$, which is
close to $\bar{R}$ of the crystalline phases of $He^4$. As is seen
from Fig.~\ref{fig1}, the distribution $|\psi(x)|^2$ at the height
of the potential barrier $U_{b} =300\,$K is similar to the bare
one ($\sin^{2}{(k_{cx}x)}$), but is more flattened. At $U_{b}
\gsim 1000\,$K, we see the appearance of two maxima located
symmetrically relative to the center of a cell. They increase with
$U_{b}.$ So, the probability density is the highest not at the
center of a cell, as is commonly accepted, but at these maxima. In
the three-dimensional case, the maxima indicate the presence,
inside a cell, of an orbit with cubic shape. The orbit depends on
the values of $a$ and $b$: at $a=1\,\mbox{\AA}$, $b=3\,\mbox{\AA}$
and $U_{b} =3000\,$K, the maxima disappear, and $|\psi(x)|^2$ is
similar to the curve of triangles in Fig.~\ref{fig1}. But such
small $a$ does not correspond to the $He^4$-$He^4$ potential. As
$a$ increases by $1\,\mbox{\AA},$ there appears a clear orbit,
like the curve of circles in Fig.~\ref{fig1}.

\begin{figure}[h]
\centerline{\includegraphics[width=85mm]{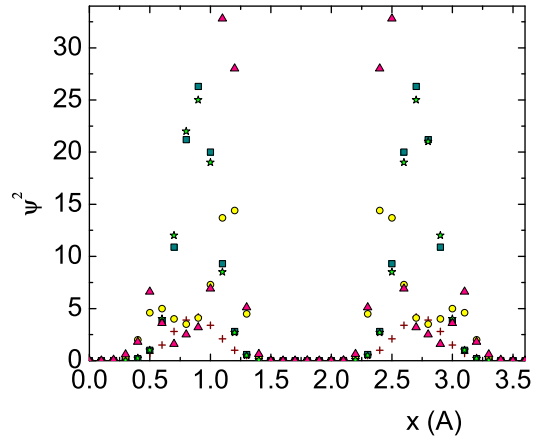}}
\caption{Function $|\psi(x)|^{2}$ for a sc crystal of $He^4$ at
high barriers $U_{b}$. Potential (\ref{2-28}) with
$a=2\,\mbox{\AA}$, $b=4\,\mbox{\AA}$,  $U_{bd} =-9\,$K and the
barriers $U_{b} =9000\,$K (crosses), $U_{b} =10^5\,$K (squares),
$U_{b} =10^6\,$K (circles) and $U_{b} =10^7\,$K (triangles). Stars
mark the curve for a crystal of krypton atoms with the same
parameters except the mass ($m=21 m_{4}$) and potential ($U_{b}
=3000\,$K, $U_{bd} =-140\,$K).
 \label{fig2}}
\end{figure}

The orbit size is approximately equal to a half of the cell size,
because the main contribution to the maxima is given by the first
term (with $q_{x}=2k_{cx}$) in sum (\ref{2-26}). The terms with
$q_{x}=4k_{cx},6k_{cx}, 8k_{cx}$ are small, and the peaks for
$q_{x}=4k_{cx}$ arise only at $U_{b} \gsim 10^6\,$K (see
Fig.~\ref{fig2}). To calculate $|\psi(x)|^2,$ it is sufficient to
take two first terms in sum (\ref{2-26}) (due to the fast decrease
of $a_{2}(q),$ as $q$ increases).

In Fig.~\ref{fig2}, we present the dependence of the orbit on the barrier height $U_{b}$.
As is seen, the orbit becomes narrow at large $U_{b}.$ Moreover, at $U_{b} =10^6\,$K and $10^7\,$K,
the second orbit appears at a distance of $\simeq 0.5\,\mbox{\AA}$ from the wall of a cell. However,
the neglected higher correlative corrections become large at large $U_{b},$ which can lead to the widening of orbits.

In Fig.~\ref{fig2}, stars show $|\psi(x)|^{2}$ for the lattice of
krypton atoms. The orbit is narrow at the barrier $U_{b}
=3000\,$K, whereas the orbit for helium atoms is wide at such
$U_{b}$. However, it is known \cite{nosanow1962,rev1} that the
ground state of the crystal of heavy inert elements is well
described under the assumption of small oscillations of atoms near
points of the lattice. Apparently, there is no contradiction in
this case, since the function $r^{2}|\psi(r)|^{2}$ at small
oscillations is characterized by a spherical orbit \cite{rev1}
with approximately the same size. Nevertheless, the orbit in
Fig.~\ref{fig2} has shape of the surface of a cube (with edge
$\bar{R}/2$), rather than a sphere, even for the function
$r^{2}|\psi(r)|^{2}$. In other words, the motion of an atom is
oscillatory only approximately. More exactly, this motion has a
wave character. This fact is unusual and means that the resonance
wave (the product of sines in (\ref{1-6})) sets the lattice of a
crystal and the motion of atoms in the cells.

Can an orbit be discovered in experiments? The scattering of light
or neutrons in a crystal will show ordinary Bragg--Wolf peaks
first of all, since the time average of the positions of an atom
is the center of a cell. But the orbit can be revealed in some
specific features of the scattering.

Let us estimate the ground-state energy $E_{0}$ of a sc crystal,
using formulas (\ref{2-10})--(\ref{2-13}). With regard for
solution (\ref{2-23}) for $S_{1}^{(\textbf{l})}(\textbf{q})$ with
$S_{1}^{(\textbf{l})}(\textbf{q}_{x})$ (\ref{2-21}) and solution
(\ref{2-9}) for $C_{l}(j),$ we obtain
\begin{equation}
 E_{0}= \tilde{E}^{b}_0 + A_{1}, \  A_{1}=  \frac{\hbar^{2}k_{c}^{2}}{2m}\left [ 1 +
 8\sum\limits_{r}\frac{a_{2}(q_{x})-3a_{2}^{2}(q_{x})}{(2a_{2}(q_{x})-1)^2}  \right ], \label{2-31}
      \end{equation}
where $r=1, 2, 3, \ldots$, $q_{x}=2\pi r/\bar{R}$,
$k_{c}^{2}=3\pi^{2}/\bar{R}^{2}$, and $\tilde{E}^{b}_0$
(\ref{2-12}) coincides with $E_{0}$ of a Bose fluid. Consider a
crystal of $He^4$ atoms with $\bar{R}=3.6\,\mbox{\AA}$,
$\hbar^{2}k_{c}^{2}/2m\approx 13.85\,$K, and the potential
(\ref{2-28}) with $a= 2\,\mbox{\AA}$, $b= 4\,\mbox{\AA}$,
$U_{bd}=-9\,$K. Using the zero approximation (\ref{2-27}) for
$a_{2}(q_{x})$ and taking the barrier $U_{b}=1000\,$K,  we obtain
$A_{1}\approx -12\,$K, $E_{0}\approx -430\,$K. For $U_{b}=200\,$K,
we have $A_{1}\approx 4\,$K, $E_{0}\approx -36\,$K, whereas
$A_{1}\approx 9\,$K, $E_{0}\approx -1\,$K for $U_{b}=81\,$K. The
last value of $E_{0}$ is close to the experimental one
\cite{rev1}. For He II, $E_{0}$ corresponds to experimental data
also for $U_{b}\sim 100\,$K (see I). For the realistic value
$U_{b}\sim 10^{3}-10^{4}\,$K, the experimental value $E_{0}\simeq
-1\,$K can be obtained with regard for correlative corrections.

The most essential and unexpected is the conclusion that the
lattice is created by a standing wave in the probability field. As
we see in Sec. 3, this wave is similar to a sound one. In other
words, the crystals have a wave nature.

 \section{State with a longitudinal acoustic phonon}
Consider a crystal with sc lattice. An optical phonons are absent
for it. We consider only longitudinal acoustic phonons. The WF of
a crystal with a single phonon can be obtained from the solution
for a one-phonon state of the Bose fluid (see I) by the following
changes: $\textbf{k}_{1}\rightarrow \textbf{k}_{c}$,
$C_{1}(q)\rightarrow C_{l}(q)$, and $S_{j}^{(1)} \rightarrow
S_{j}^{(\textbf{l})}$. We obtain
  \begin{equation}
    \Psi_{\textbf{k}}(\textbf{r}_1,\ldots ,\textbf{r}_N) =
  \psi_{\textbf{k}}\Psi_0,
  \label{3-1}     \end{equation}
  \begin{equation}
 \psi_{\textbf{k}}  = \psi^{0}_{\textbf{k}} + 7\,\mbox{permutations},
       \label{3-2} \end{equation}
 \begin{eqnarray}
&& \psi^{0}_{\textbf{k}}  =  \psi^{b}_{\textbf{k}} + b_{0}(\textbf{k}) +
 \sum\limits_{\textbf{q} \neq 0, -\textbf{k}}^{(2\pi)}
  Q_{1}(\textbf{q}, \textbf{k})\rho_{-\textbf{q}-\textbf{k}}  \nonumber\\
 & + & \sum\limits_{\textbf{q}, \textbf{q}_{1}\neq 0}^{\textbf{q}+\textbf{q}_{1}+
 \textbf{k} \not= 0}
  \frac{Q_{2}(\textbf{q},\textbf{q}_{1},\textbf{k})}{\sqrt{N}}
 \rho_{\textbf{q}_{1}}\rho_{-\textbf{q}-\textbf{q}_{1}-\textbf{k}}
  \label{3-3} \\
 & + & \sum\limits_{\textbf{q}, \textbf{q}_{1}, \textbf{q}_{2}\neq 0}^{\textbf{q}+\textbf{q}_{1}+\textbf{q}_{2}
 +\textbf{k} \not= 0}
  \frac{Q_{3}(\textbf{q},\textbf{q}_{1},\textbf{q}_{2},\textbf{k})}{N}
 \rho_{\textbf{q}_{1}}\rho_{\textbf{q}_{2}}\rho_{-\textbf{q}-\textbf{q}_{1}-\textbf{q}_{2}-\textbf{k}}
 +\ldots ,
      \nonumber  \end{eqnarray}
        \begin{eqnarray}
 & &  \psi^{b}_{\textbf{k}} = \rho_{-\textbf{k}} +
 \sum\limits_{\textbf{k}_{2}\neq 0, -\textbf{k}}^{(\pi)}
  \frac{b_{2}(\textbf{k}_{2},\textbf{k})}{\sqrt{N}}
 \rho_{\textbf{k}_{2}}\rho_{-\textbf{k}_{2}-\textbf{k}}  \label{3-4} \\
 & + & \sum\limits_{\textbf{k}_{2},\textbf{k}_{3}\neq 0}^{(\pi)\,\textbf{k}_{2}+
 \textbf{k}_{3}+\textbf{k} \not= 0}
  \frac{b_{3}(\textbf{k}_{2},\textbf{k}_{3},\textbf{k})}{N}
 \rho_{\textbf{k}_{2}}\rho_{\textbf{k}_{3}}\rho_{-\textbf{k}_{2}-\textbf{k}_{3}-\textbf{k}}
 +\ldots,
       \nonumber \end{eqnarray}
 where the permutation means $\psi^{0}_{\textbf{k}}$ with the different sign of one or several components of the vector
$\textbf{k}$, the vector $\textbf{q}$ is quantized like $2\pi
j/L$, and the vectors  $\textbf{k}_{j}$, $\textbf{q}_{j},$ and $\textbf{k}$ are quantized like $\pi j/L$.

 Solution (\ref{3-2}) describes a three-dimensional standing wave decaying into eight counter traveling waves.

The energy of a phonon $E(k)$ and the functions  $b_{j}$,  $Q_{j}$ satisfy the equations
\begin{eqnarray}
 &\epsilon(k) & =  \epsilon_{0}(k)-\frac{1}{N}\sum\limits_{\textbf{k}_{2}\neq 0}^{(\pi)}
 b_{2}(\textbf{k}_{2}, \textbf{k})2\textbf{k}_{2}(\textbf{k}_{2}+\textbf{k})
  \nonumber\\
  &-&\frac{1}{N}\sum\limits_{\textbf{k}_{2}\neq 0}^{(\pi)}
 6k^{2}_{2} b_{3}(\textbf{k}_{2}, -\textbf{k}_{2}, \textbf{k})
  \nonumber\\
  & - & \frac{2}{\sqrt{N}}\sum\limits_{\textbf{q}\neq 0}^{(2\pi)} Q_{1}(\textbf{q},\textbf{k})(\textbf{k}+\textbf{q})
   \nonumber \\ &\times&
 \left [ \textbf{q}S_{1}^{(\textbf{l})}(-\textbf{q})+2(\textbf{k}+\textbf{q})S^{(\textbf{l})}_{2}(-\textbf{q}, -\textbf{k})
  \right ] \nonumber\\
     & - & \frac{4}{\sqrt{N}}\sum\limits_{\textbf{q}\neq 0}^{(2\pi)} q^{2}Q_{2}(-\textbf{q},\textbf{q},\textbf{k})
S_{1}^{(\textbf{l})}(\textbf{q}) \label{3-5}\\
   & - & \sum\limits_{q_{x}\neq 0}^{(2\pi)} 2k_{cx}(k_{x}+q_{x})i C_{l_{x}}(-q_{x}) Q_{1}(\textbf{q}_{x},\textbf{k})
   \nonumber\\
   & + & \sum\limits_{q_{x}\neq 0}^{(2\pi)} 4k_{cx}q_{x}i C_{l_{x}}(q_{x}) Q_{2}(-\textbf{q}_{x},\textbf{q}_{x},\textbf{k})
   + (x\rightarrow y, z),
     \nonumber   \end{eqnarray}
\begin{eqnarray}
&& Q_{1}(\textbf{q},\textbf{k})\left
[\epsilon(k)-\epsilon_{0}(\textbf{k}+\textbf{q})
 \right ]=  \nonumber\\
 & = & 2ik_{cx}C_{l_{x}}(q_{x})\delta_{\textbf{q},\textbf{q}_{x}}\left [
 -k_{x}+2q_{x}b_{2}(\textbf{q}, \textbf{k})
  \right ] \nonumber\\
  &+& \frac{2S_{1}^{(\textbf{l})}(\textbf{q})}{\sqrt{N}}[\textbf{q}\textbf{k}-2q^{2}b_{2}(\textbf{q}, \textbf{k})]
  -\frac{4k^2}{\sqrt{N}}S_{2}^{(\textbf{l})}(\textbf{q},-\textbf{q}-\textbf{k})
  \nonumber\\ &-& \frac{1}{N}\sum\limits_{\textbf{q}_{1}\neq 0}^{(\pi)} \left \{ 2\textbf{q}_{1}(\textbf{q}_{1}+\textbf{q}+
  \textbf{k})\right. \nonumber\\ &\times &
  [Q_{2}(\textbf{q},\textbf{q}_{1},\textbf{k})+Q_{1}(\textbf{q}+\textbf{q}_{1},\textbf{k})\sqrt{N}S_{1}^{(\textbf{l})}(-\textbf{q}_{1}) ] \nonumber\\
  &+& 6q_{1}^{2}Q_{3}(\textbf{q},\textbf{q}_{1},-\textbf{q}_{1},\textbf{k}) \nonumber\\
  &+& 4q_{1}^{2}Q_{2}(\textbf{q}-\textbf{q}_{1},\textbf{q}_{1},\textbf{k})\sqrt{N}S_{1}^{(\textbf{l})}(\textbf{q}_{1}) \nonumber\\
  &+& \left. 4(\textbf{q}_{1}+\textbf{q}+ \textbf{k})^{2}Q_{1}(\textbf{q}+
  \textbf{q}_{1},\textbf{k})\sqrt{N}S_{2}^{(\textbf{l})}(-\textbf{q}_{1},-\textbf{q}-\textbf{k}) \right \} \nonumber\\
  &+& \sum\limits_{p_{x}\neq 0}^{(2\pi)} 2k_{cx}iC_{l_{x}}(p_{x})
  \left [(-k_{x}-q_{x}+p_{x})Q_{1}(\textbf{q}-\textbf{p}_{x},\textbf{k})\right. \nonumber \\
  &+& \left. 2p_{x}
  Q_{2}(\textbf{q}-\textbf{p}_{x},\textbf{p}_{x},\textbf{k}) \right ]+(x\rightarrow y, z),
       \label{3-6} \end{eqnarray}
        \begin{eqnarray}
 && b_{0}(\textbf{k})\epsilon(k)  =  -2k^{2}S_{1}^{(\textbf{l})}(-\textbf{k})\delta_{\textbf{k},\textbf{k}^{e}}  \nonumber\\
   & - & 2k_{cx}k_{x}iC_{l_{x}}(-k_{x})\sqrt{N}\delta_{\textbf{k},\textbf{k}_{x}^{e}}
   \nonumber\\
    & - & \frac{2}{\sqrt{N}}\sum\limits_{\textbf{q}\neq 0}^{(\pi)}q^{2}Q_{2}(-\textbf{k},\textbf{q},\textbf{k})\delta_{\textbf{k},\textbf{k}^{e}}
 \nonumber\\ &-& \sum\limits_{\textbf{q}\neq 0, -\textbf{k}}^{(2\pi)}2(\textbf{k}+\textbf{q})^{2}Q_{1}(\textbf{q},\textbf{k})
  S_{1}^{(\textbf{l})}(-\textbf{k}-\textbf{q})\delta_{\textbf{k},\textbf{k}^{e}} \nonumber\\
    & + & \sum\limits_{q_{x}\neq 0}^{(2\pi)}2\sqrt{N}k_{cx}q_{x}iC_{l_{x}}(q_{x})Q_{1}(-\textbf{k}-\textbf{q}_{x},\textbf{k})\delta_{\textbf{k},\textbf{k}^{e}}
  \nonumber\\ &+&  (x\rightarrow y, z),
       \label{3-7} \end{eqnarray}
\begin{eqnarray}
&& b_{2}(\textbf{k}_{2}, \textbf{k})\left
[\epsilon(k)-\epsilon_{0}(\textbf{k}_{2})-\epsilon_{0}(\textbf{k}+\textbf{k}_{2})
 \right ]=  \nonumber\\
 & = & \textbf{k}\textbf{k}_{2}a_{2}(\textbf{k}_{2})-\textbf{k}(\textbf{k}+\textbf{k}_{2})
 a_{2}(\textbf{k}+\textbf{k}_{2})-k^{2}a_{3}(\textbf{k}, \textbf{k}_{2}) \nonumber\\
 &-& \frac{2}{\sqrt{N}}\sum\limits_{\textbf{k}_{3}\neq 0}^{(\pi)} \left \{ \frac{3\textbf{k}_{3}(\textbf{k}_{2}+\textbf{k}_{3}+
  \textbf{k})}{\sqrt{N}}b_{3}(\textbf{k}_{2},\textbf{k}_{3},\textbf{k})\right. \nonumber\\ &+ &
  Q_{1}(\textbf{k}_{3},\textbf{k})(\textbf{k}_{3}+\textbf{k})\left [2(\textbf{k}_{3}-\textbf{k}_{2})S_{2}^{(\textbf{l})}(-\textbf{k}_{3},\textbf{k}_{2})
  \right. \nonumber\\
  &+&\left. 3(\textbf{k}_{3}+\textbf{k})S_{3}^{(\textbf{l})}(-\textbf{k}_{3},\textbf{k}_{2},-\textbf{k}_{2}-\textbf{k})\right ] \nonumber\\
  &+& 2Q_{2}(-\textbf{k}_{3},\textbf{k}_{2}+\textbf{k}_{3},\textbf{k})(\textbf{k}_{2}+\textbf{k}_{3}) \nonumber\\
  &\times& \left [\textbf{k}_{3}S_{1}^{(\textbf{l})}(\textbf{k}_{3})+2(\textbf{k}_{2}+\textbf{k}_{3})S_{2}^{(\textbf{l})}(\textbf{k}_{3},\textbf{k}_{2}) \right ]
   \nonumber\\ &+& \left. 3k_{3}^{2}S_{1}^{(\textbf{l})}(\textbf{k}_{3})Q_{3}(-\textbf{k}_{3},\textbf{k}_{2},-\textbf{k}_{2}-\textbf{k},\textbf{k}) \right \}
  \nonumber\\ &+& \sum\limits_{p_{x}\neq 0}^{(2\pi)} 2k_{cx}iC_{l_{x}}(p_{x})
  \left [3p_{x}Q_{3}(-\textbf{p}_{x},\textbf{k}_{2},-\textbf{k}_{2}-\textbf{k},\textbf{k})\right. \nonumber \\
  &+& \left. 2(k_{2x}+p_{x})
  Q_{2}(-\textbf{p}_{x},\textbf{k}_{2}+\textbf{p}_{x},\textbf{k}) \right ],
 \nonumber\\ &+& \mbox{higher corrections} +(x\rightarrow y, z),
       \label{3-8} \end{eqnarray}
where $\epsilon(k)=2mE(k)/\hbar^2$, $\textbf{k}^{e}$ is the wave vector with all even components (i.e., they are multiple to $2\pi/L$),
$\textbf{q}_{x}=q_{x}\textbf{i}_{x},$ and $\textbf{p}_{x}=p_{x}\textbf{i}_{x}$. By
$(x\rightarrow y, z),$ we denote
the same terms as one with the separated $x$-component, but with the changes $x\rightarrow y$ and $x\rightarrow z$.
All terms of Eqs. (\ref{3-5})--(\ref{3-8}), except for the first one, contain a product of two wave vectors
(for example, $q^2$,  $k_{cx}k_{x},$ or
$(\textbf{k}_{3}+\textbf{k})(\textbf{k}_{3}-\textbf{k}_{2})$). These wave vectors must be nonzero.

Some of these equations contain the sums, where the first argument
of the functions $Q_{l}$ or $S_{j}^{(\textbf{l})}$ has components
multiple to $\pi/L$. In this case, we set $Q_{l}=0$,
$S_{j}^{(\textbf{l})}=0$, because, by the definition of these
functions, the components of the first argument are multiple to
$2\pi/L$.

Equations (\ref{3-5})--(\ref{3-8}) are written in the approximation
of ``two sums in the wave vector'', at which the series contain the
functions $a_{2}$, $a_{3}$, $b_{2}$, $b_{3}$, $S^{(1)}_{j\leq 3}$, and $Q_{l\leq 3}$ and
do not include the corrections $a_{j\geq 4}$, $b_{j\geq 4}$, $S^{(1)}_{j\geq 4},$
and $Q_{l\geq 4}$.

It follows from Eq. (\ref{3-6}) that the solution for the function
$Q_{1}(\textbf{q},\textbf{k})$ has a ``resonance'' form analogous
to (\ref{2-20}). We restrict ourselves to the one-dimensional
approximation:
\begin{equation}
Q_{1}^{(\textbf{l})}(\textbf{q},\textbf{k}) \simeq
Q_{1}^{(\textbf{l})}(\textbf{q},\textbf{k})\left \{\delta_{\textbf{q},\textbf{q}^{res}_{x}}+
\delta_{\textbf{q},\textbf{q}^{res}_{y}} +
\delta_{\textbf{q},\textbf{q}^{res}_{z}}  \right \}.
      \label{3-9} \end{equation}
In the zero approximation without regard for the sums in Eq.
(\ref{3-6}), we have
\begin{eqnarray}
 Q_{1}(\textbf{q}_{x},\textbf{k})&=&  
 \left [2ik_{cx}C_{l_{x}}(q_{x}) - 2q_{x}S_{1}^{(\textbf{l})}(\textbf{q}_{x})/\sqrt{N}\right ]\nonumber \\
 & \times &\frac{-k_{x}+2q_{x}b_{2}(\textbf{q}_{x},
  \textbf{k})}{\epsilon(k)-\epsilon_{0}(\textbf{k}+\textbf{q}_{x})} \label{3-10}  \\
  &\approx & \frac{i2k_{cx}C_{l_{x}}(q_{x})}{1-2a_{2}(q_{x})}
 \times\frac{-k_{x}+2q_{x}b_{2}(\textbf{q}_{x},
  \textbf{k})}{\epsilon(k)-\epsilon_{0}(\textbf{k}+\textbf{q}_{x})}.
       \nonumber \end{eqnarray}
The analogous relations can be given for $Q_{1}(\textbf{q}_{y},\textbf{k}),$ and $Q_{1}(\textbf{q}_{z},\textbf{k})$.

Equations (\ref{3-5})--(\ref{3-8}) are very complicated. By
setting $Q_{l}=0$ in them, we obtain the equations for a Bose
fluid, namely, for $b_{2}$ and the dispersion curve $E(k)$.
 Thus, the equations for a crystal are
the equations for a fluid plus some additional anisotropic corrections.

Let us consider dispersion curves. First,  the value of $b_{0}$
does not influence $E(k)$. In the simplest approximation with
regard for the anisotropy, the dispersion curves are determined by
the formula
\begin{eqnarray}
 &\epsilon(\textbf{k}) & =  \epsilon_{0}(k)- 
 \frac{2}{\sqrt{N}}\sum\limits_{\textbf{q}\neq 0}^{(2\pi)} Q_{1}(\textbf{q},\textbf{k})(\textbf{k}+\textbf{q})
     \textbf{q}S_{1}^{(\textbf{l})}(-\textbf{q})     
   \nonumber\\
   & - & \sum\limits_{q_{x}\neq 0}^{(2\pi)} 2k_{cx}(k_{x}+q_{x})i C_{l_{x}}(-q_{x}) Q_{1}(\textbf{q}_{x},\textbf{k})
   \nonumber\\ &+& (x\rightarrow y, z).
       \label{3-11} \end{eqnarray}
With regard for solutions (\ref{2-9}), (\ref{2-21}), (\ref{2-23}),  (\ref{3-9}), and (\ref{3-10}) in the approximation $b_{2}=0,$ we obtain
\begin{equation}
 \epsilon(\textbf{k})  =  \epsilon_{0}(k)+\epsilon^{cr}_{x}(\textbf{k})+\epsilon^{cr}_{y}(\textbf{k})+\epsilon^{cr}_{z}(\textbf{k}),
        \label{3-12} \end{equation}
    \begin{equation}
 \epsilon^{cr}_{x}(\textbf{k}) =   - \sum\limits_{q_{x}=q_{x}^{res}}\frac{4k^{2}_{cx}}{(1-2a_{2}(q_{x}))^2}
 \frac{k_{x}(k_{x}+q_{x})}{\epsilon(\textbf{k})-\epsilon_{0}(\textbf{k}+\textbf{q}_{x})},
       \label{3-13} \end{equation}
where the sum is taken over the resonance values of $q_{x}^{res} = \pm 2k_{cx}, \pm 4k_{cx}, \pm 6k_{cx}, \ldots$.
For a sc crystal,
\begin{equation}
 E(\textbf{k})  =  \frac{\hbar^{2}\epsilon_{0}(\textbf{k})}{2m}+
 j\frac{\hbar^{2}\epsilon^{cr}_{x}(\textbf{k})}{2m},
        \label{3-14} \end{equation}
where $j$=1, 2, and 3 for directions (1,0,0), (1,1,0), and (1,1,1). Since the values of  $\epsilon^{cr}_{x}(\textbf{k})$
and $\epsilon_{0}(k)$ are of the same order, the dispersion curves are different for different directions.
In other words, the spectrum of longitudinal acoustic phonons is anisotropic,
what was observed in experiments.

\begin{figure}[h]
\centerline{\includegraphics[width=85mm]{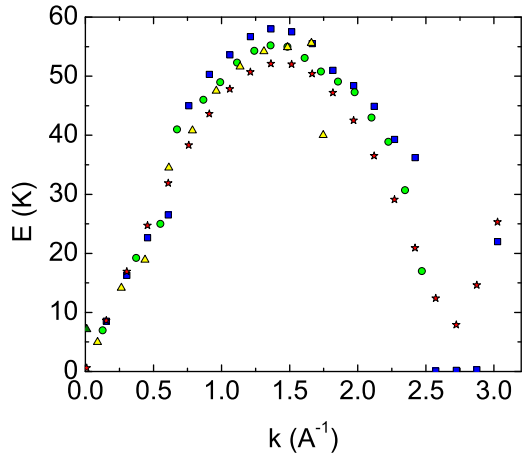}}
\caption{Dispersion curves $E(\textbf{k})$ (\ref{3-14}) with
$\epsilon^{cr}_{x}$ decreased by two orders of magnitude for
directions (1,0,0) (triangles), (1,1,0) (circles), and (1,1,1)
(squares) for a sc crystal of $He^4$ atoms with
$\bar{R}=3.6\,\mbox{\AA}$ and potential (\ref{2-28}) with $a=
2\,\mbox{\AA}$, $b= 4\,\mbox{\AA}$,  $U_{bd}=-9\,$K and $U_{b}
=3000\,$K.
 Stars show the zero approximation $E(\textbf{k}) = \hbar^{2}\epsilon_{0}(k)/2m$.
 For the comparison with experiments, the dispersion curves are drawn to $k^{max}$ of the bcc lattice.
 \label{fig3}}
\end{figure}
Due to a large value of $\epsilon^{cr}_{x},$ the method of
iterations does not converge, and the more exact numerical methods
require much time for the analysis. To demonstrate the influence
of the correction $\epsilon^{cr}_{x}$, we decrease it by two
orders of magnitude so that the method of iterations can be
applied. In Fig. 3, we present the dispersion curves calculated by
formula (\ref{3-14}). The curves are different for different
directions and are similar to the observed ones \cite{rev2} for a
bcc crystal of $He^4$.

The qualitative behavior of the dispersion curves can be understood, by using the simple formula
\begin{equation}
 E(\textbf{k})  \approx  \frac{\hbar^{2}\epsilon_{0}(\textbf{k})}{2m} =
 \frac{\hbar^{2}k^{2}}{2m}(1-2a_{2}(\textbf{k})).
       \label{3-15} \end{equation}
With regard for approximation (\ref{2-27}) for $a_{2}(\textbf{k}),$ this formula becomes
\begin{equation}
 E(k) \approx \sqrt{\left (\frac{\hbar^{2} k^2}{2m}\right )^{2} +
  \frac{n\nu_{3}(k)}{8}\frac{\hbar^2 k^2}{m}}.
      \label{3-16} \end{equation}
It is the well-known Bogolyubov formula with the additional
multiplier $1/8$ arising due to the boundaries (see I). Indeed,
for each of the directions (1,0,0), (1,1,0), and (1,1,1), we need
to continue curve (\ref{3-16}) to  $k^{max}$ for a given
direction. We obtained no anisotropy, but the general shape of the
curve is proper: the phonon curve has no minimum for direction
(1,0,0), a slightly pronounced minimum for (1,1,0), and a clear
minimum for (1,1,1). This corresponds to experiments \cite{rev2} for the bcc
lattice of $He^4$. The minimum arise due to a
displacement of $k^{max}$ to the side of large $k$ and to the
presence of a minimum on the liquid-like curve $E(\textbf{k})  =
\hbar^{2}\epsilon_{0}(k)/2m$. To  describe the curve more exactly
and to obtain the anisotropy, we need to consider the next
corrections to the equations.

It is of interest that the simple formula (\ref{3-16}) describes
the curves qualitatively correctly. This formula is universal and
is valid for a gas, a liquid, and a crystal, since these three
states of substance are described by WFs with the same structure
(\ref{1-6}), (\ref{3-1})--(\ref{3-4}) (the difference is only in
the values of $\textbf{k}_{\textbf{l}}$).

The theoretical dispersion curves for a crystal and He II (see I)
correspond to the experiment at $U_{b} \sim 10^3\,$K. For $E_{0},$
the agreement with experiment for a liquid and a crystal holds at
$U_{b} \sim 10^2\,$K (see Sec. 2). However, it is assumed that the
best potential for $He^4$ atoms is the Aziz potential \cite{aziz}
with much more larger value of $U_{b}$, about $2\times 10^6\,$K.
This disagreement is related to the neglect of higher corrections
or (more probably) to the fact that the potential at small
distances has only an effective meaning. Therefore, the values of
$U_{b}$ can be very different for different processes. In some
models, the agreement with experiment was attained with the Aziz
potential.

In Fig.~\ref{fig3}, the dispersion curves are broken at $\textbf{k}^{max}$ equal to a half of the minimum vector $\textbf{g}_{0}$ of the reciprocal lattice
(for directions (1,0,0), (1,1,0), and (1,1,1) of a sc crystal,
$\textbf{g}_{0}=\pi(1/\bar{R}, 0, 0), \pi(1/\bar{R}, 1/\bar{R}, 0)$ and $\pi(1/\bar{R}, 1/\bar{R}, 1/\bar{R})$,
whereas $\textbf{g}_{0}$ is twice larger for the bcc lattice), since the dispersion curves for crystals are periodic with period $\textbf{g}$:
\begin{equation}
E(\textbf{k}+\textbf{g})=E(\textbf{k}).
      \label{3-17} \end{equation}
The proof of this fact for one-particle WFs \cite{kittel1963} can
be easily generalized to our case of $N$-particle WFs. However, it
is true for the cyclic BCs. For a
three-dimensional crystal, the BCs are quite
different and are close to zero ones. However, the majority of
quasiparticles are localized wave packets (this is supported by
the fact that the theory of transfer was constructed for wave
packets and agrees with experiment). For such packets, the
translational invariance with the period of a lattice holds, and
the conclusions of the theorem are valid. As for the standing
waves (\ref{3-1})--(\ref{3-4}), they are the sum of eight
traveling waves, from which the wave packets can be constructed.

We note that the density of phonon states at the quantization of $\textbf{k}$ by law (\ref{2-2}) is the same as that
for cyclic BCs.

In the literature \cite{nosanow1967,rev1,glyde1976}, the dispersion curves of crystals are usually calculated in the approximation of small oscillations (\ref{1-1}).

\section{The condensate}
In Ref.~\onlinecite{penronz1956} and recently  \cite{ceperley2006}, it
was shown that the ideal crystal in the ground state contains no
condensate atoms with $\textbf{k}=0$. However, the condensate is
possible in the presence of vacancies and other defects
\cite{andreev1969,spigs2006,superglass,grain,disloc-net,rota2011}.
The condensate can be a key factor for the explanation of NCIM.
But the experiment does not confirm \cite{diallo2011} a condensate
of atoms with $\textbf{k}=0$.

It follows from $\Psi_{0}$ (\ref{1-6}) that the ideal sc crystal
does not possess off-diagonal long-range order, but it have a
condensate of atoms with $\textbf{k}=\textbf{k}_{\textbf{l}}$. It
is easy to verify if we switch-off the interatomic interaction: in
this case, the exponents in $\Psi_{0}$ (\ref{1-6}) become equal to
1, and we have the product of sines with
$\textbf{k}=\textbf{k}_{\textbf{l}}$, i.e., all atoms are in the
condensate with $\textbf{k}=\textbf{k}_{\textbf{l}}$. If the
interaction is switched-on, the condensate is exhausted. For a
fluid, the condensate is also determined by a factor before the
exponential function: under cyclic BCs, this
factor ($\prod\limits_{j}
e^{i\textbf{p}\textbf{r}_{j}}|_{\textbf{p}=0} = 1$) generates the
condensate of atoms with $\textbf{p}=0$. For the fluid in a
vessel, the factor is the product of sines with
$\textbf{k}=\textbf{k}_{1}$ (see I), and the condensate is on the
levels with $\textbf{k}=\textbf{k}_{1}, 3\textbf{k}_{1},
5\textbf{k}_{1}, \ldots$ \cite{zero-cond}.

Let us calculate the condensate for the ground state of a sc
crystal. Under zero BCs, the condensate is
determined by the formula \cite{zero-cond}
\begin{eqnarray}
\frac{N_{\textbf{k}_{\textbf{n}}}}{N}&=&\frac{8}{V^2}\int\limits_{0}^{L_{x}, L_{y}, L_{z}} d\textbf{r}_{1} d\textbf{r}_{2}
\rho(\textbf{r}_{1},\textbf{r}_{2})\sin{(k_{n_{x}}x_{1})}\sin{(k_{n_{x}}x_{2})}\nonumber \\
&\times &\sin{(k_{n_{y}}y_{1})}
\sin{(k_{n_{y}}y_{2})}\sin{(k_{n_{z}}z_{1})}
\sin{(k_{n_{z}}z_{2})},
\label{c-1}  \end{eqnarray}
\begin{equation}
\rho(\textbf{r}_{a},\textbf{r}_{b}) = V\int d\acute{\textbf{r}}
\Psi_{0}^{*}(\textbf{r}_{a},\acute{\textbf{r}})
\Psi_{0}(\textbf{r}_{b},\acute{\textbf{r}}), \label{c-2}
\end{equation} where  $n_{x}, n_{y}, n_{z}$  are integers,
$\textbf{n}=(n_{x}, n_{y}, n_{z})$, and $\acute{\textbf{r}}$ marks
a collection of vectors $\textbf{r}_{2},\ldots,\textbf{r}_{N}$.
Relation (\ref{1-6}) yields
\begin{eqnarray}
\rho(\textbf{r}_{a},\textbf{r}_{b})& =&
\sin{(k_{l_{x}}x_{a})}\sin{(k_{l_{y}}y_{a})} \sin{(k_{l_{z}}z_{a})}\nonumber \\
&\times &\sin{(k_{l_{x}}x_{b})}\sin{(k_{l_{y}}y_{b})} \sin{(k_{l_{z}}z_{b})}
\nonumber \\ &\times &|A|^{2}V\int d\acute{\textbf{r}}\left [
\exp{\left \{(S_{w}^{(\textbf{l})}(\textbf{r}_{a},\acute{\textbf{r}}))^{*}\right \} }\right. \label{c-3} \\
&\times &  \exp{\left \{S_{w}^{(\textbf{l})}(\textbf{r}_{b},\acute{\textbf{r}})
+\tilde{S}_{b}(\textbf{r}_{a},\acute{\textbf{r}})+\tilde{S}_{b}(\textbf{r}_{b},\acute{\textbf{r}})\right \}} \nonumber \\
&\times &\left. \prod\limits_{j=2}^{N}\{\sin^{2}{(k_{l_{x}}x_{j})}\sin^{2}{(k_{l_{y}}y_{j})}
 \sin^{2}{(k_{l_{z}}z_{j})}\}\right ].
  \nonumber \end{eqnarray}
In $S_{w}^{(\textbf{l})}$ (\ref{2-3}), we consider only the one-particle part $S_{1}^{(\textbf{l})}$. Then
 \begin{eqnarray}
\rho(\textbf{r}_{a},\textbf{r}_{b}) &=&\rho^{c}_{\infty}
\sin{(k_{l_{x}}x_{a})}\sin{(k_{l_{y}}y_{a})} \sin{(k_{l_{z}}z_{a})}\nonumber \\
&\times &\sin{(k_{l_{x}}x_{b})}\sin{(k_{l_{y}}y_{b})} \sin{(k_{l_{z}}z_{b})}
\nonumber \\ &\times &
\exp{\left \{\sum\limits_{\textbf{q}_{x}=\textbf{q}^{res}_{x}} R_{\rho}(\textbf{q}_{x})
+(x\rightarrow y, z)\right \}}.
 \label{c-4}  \end{eqnarray}
 \begin{equation}
R_{\rho}(\textbf{q}_{x})=\frac{1}{\sqrt{N}}\left ( (S_{1}^{(\textbf{l})}(\textbf{q}_{x}))^{*}e^{-iq_{x}x_{a}}+
S_{1}^{(\textbf{l})}(\textbf{q}_{x})e^{iq_{x}x_{b}}\right ).
 \label{c-4b}  \end{equation}
We use approximation (\ref{2-23}), (\ref{2-21}). Then $S_{1}^{(\textbf{l})}$ is real, and we can replace
$e^{-iq_{x}x}\rightarrow \cos{q_{x}x}$ and can sum only over  $q_{x}> 0$. We obtain
\begin{equation}
\rho(\textbf{r}_{a},\textbf{r}_{b}) =\rho^{c}_{\infty}f_{x}(x_{a})f_{x}(x_{b})f_{y}(y_{a})f_{y}(y_{b})f_{z}(z_{a})f_{z}(z_{b}),
 \label{c-5}  \end{equation}
 \begin{equation}
f_{x}(x) =\sin{(k_{l_{x}}x)}\exp{\left \{\frac{2}{\sqrt{N}}\sum\limits_{q^{res}_{x}>0}S_{1}^{(\textbf{l})}(\textbf{q}^{res}_{x})\cos{(q^{res}_{x}x)}
\right \}}.
 \label{c-6}  \end{equation}
Since $q^{res}_{x}=2k_{l_{x}}, 4k_{l_{x}}, 6k_{l_{x}}, \ldots$, we have
 \begin{equation}
f_{x}(x) =\sin{(k_{l_{x}}x)}e^{\left \{g_{2}\cos{(2k_{l_{x}}x)}+g_{4}\cos{(4k_{l_{x}}x)}+\ldots \right \}},
 \label{c-7}  \end{equation}
where
$g_{j}=2S_{1}^{(\textbf{l})}(jk_{l_{x}}\textbf{i}_{x})/\sqrt{N}\sim
1$. Function (\ref{c-7}) coincides with (\ref{2-24}),
(\ref{2-26}). Substituting (\ref{c-5}) and (\ref{c-7}) in
(\ref{c-1}) and expanding (\ref{c-7}) in a series, we obtain that
the \textit{condensate levels} with $N_{\textbf{k}}\sim N$
correspond to the wave vectors $\textbf{k}=((1+2j_{x})k_{l_{x}},
(1+2j_{y})k_{l_{y}}, (1+2j_{z})k_{l_{z}})$ with $j_{x}, j_{y},
j_{z}=0, 1, 2, 3, \ldots $; i.e., to the vector
$\textbf{k}_{\textbf{l}}=\textbf{k}_{c}$ and to larger vectors
with odd multiple components.

The distribution of atoms over levels depends on the barrier
height $U_{b}$, here we have the interesting picture. The
numerical calculation for $He^4$ atoms with
$\bar{R}=3.6\,\mbox{\AA}$ and potential (\ref{2-28}) with $a=
2\,\mbox{\AA}$, $b= 4\,\mbox{\AA}$, $U_{bd}=-9\,$K, and realistic
values $500\,K\lsim U_{b}\lsim 5000\,$K gives
 $g_{2}\sim 1$, $g_{4}\sim -0.1$, and the subsequent $g_{j}$ are small.
With such $g_{j},$ we obtain $N_{\textbf{k}=\textbf{k}_{c}}\approx
(1-g_{2}/2+g_{2}^{2}/4)^{6}\rho^{c}_{\infty} N/8$. For the higher
levels, we have: $N_{\textbf{k}=3\textbf{k}_{c}}\approx
(g_{2}/2-g_{4}/2)^{6}\rho^{c}_{\infty} N/8$,
$N_{\textbf{k}=5\textbf{k}_{c}}\approx
(g_{2}^{2}/8+g_{4}/2)^{6}\rho^{c}_{\infty} N/8$,
$N_{\textbf{k}=7\textbf{k}_{c}} \approx
(g_{2}g_{4}/4)^{6}\rho^{c}_{\infty} N/8$, etc. Hence, the levels
with $\textbf{k}=\textbf{k}_{c}, 3\textbf{k}_{c}$ and the
intermediate ones with $\textbf{k}=(k_{cx}, 3k_{cy}, k_{cz})$,
$(k_{cx}, 3k_{cy}, 3k_{cz})$ (and with permutations) are filled.
The rest levels are almost empty. Thus, we have 8 levels with
$N_{\textbf{k}}\sim (g_{2}/2)^{6}\rho^{c}_{\infty} N/8\sim
\rho^{c}_{\infty} N/8^3$ atoms on each of them, in the sum
$N_{c}\sim \rho^{c}_{\infty} N/64$. For comparison, $
N_{\textbf{k}_{1}}\approx \rho_{\infty} N/2$ for a liquid
\cite{zero-cond}. It is easy to verify that
\begin{equation}
\rho^{c}_{\infty}\sim \rho^{l}_{\infty}/P^3, \quad P=\int\limits_{0}^{L_{x}}\frac{dx}{L_{x}}[f_{x}(x)]^{2}.
 \label{c-ro}  \end{equation}
With the obtained $g_{j},$ we have $P\simeq
(1-g_{2}+g_{2}^{2})/2\simeq 1/2$, i.e., $\rho^{c}_{\infty}\sim
8\rho^{l}_{\infty}$. The value of $\rho^{l}_{\infty}$ is given by
the formula for a fluid. Let us consider the ``crystal''
correction $A_{2}(\textbf{k})$ (\ref{2-15}) in Eq. (\ref{2-13})
for $a_{2}(\textbf{k}).$ Then, for $U_{b}\sim 10^{2}$-$10^{4}\,$K,
the values of $a_{2}(\textbf{k})$ decrease twice (in modulus) on
the average in the significant interval of $k$ (where $a_{2}$ is
not small). In the zero approximation \cite{vak1990},
$\rho^{l}_{\infty}=\exp{[-\frac{1}{N}\sum\limits_{\textbf{k}\neq
0}\frac{a_{2}^{2}(\textbf{k})}{1-2a_{2}(\textbf{k})}]}$.  We know
from experiment that $\rho_{\infty} \approx 0.07$ for He II. At
twice less $a_{2}(\textbf{k}),$ we obtain for $He^4$ atoms
$\rho^{l}_{\infty}\simeq 1/3$  and $\rho^{c}_{\infty}\sim 8/3$.
Hence, the condensate levels of a sc solid $He^4$ contain
$N_{c}\sim  0.04N$ atoms. This is only a rough estimate.

At $U_{b}\gsim 10^{4}$-$10^{5}\,$K, we have  $g_{2}\sim 1$, $g_{4}$ takes values from -2 down to -10,  $g_{j\geq 6}\approx 0$;
the levels with $\textbf{k}=3\textbf{k}_{c}, 5\textbf{k}_{c}, 7\textbf{k}_{c}$
and the intermediate ones are filled.
At $U_{b}\gsim 10^6\,$K, the values of $g_{6}$, $g_{8},$ and $g_{10}$ become large, which generates the condensates with $\textbf{k}= 9\textbf{k}_{c},
11\textbf{k}_{c}, 13\textbf{k}_{c}$.

As is seen, the condensate structure in a crystal is similar to
that of a fluid in a vessel \cite{zero-cond}, but with the vector
$\textbf{k}_{c}$ instead of $\textbf{k}_{1}$ and with a different
distribution over levels.  Within our method, it is impossible to
calculate the condensate with high accuracy, because the
significant corrections were omitted in almost all equations. In
this case, the Monte-Carlo method can be efficient
\cite{ceperley2006}.

The structure of $\Psi_{0}$ (\ref{1-6}) helps us to imagine the
condensate: $\Psi_{0}$ contains $N$ identical standing waves (the
product of sines) resting on the walls by their wings. These waves
form a single resonance classical wave, which modulates the motion
of atoms. Therefore, many atoms are characterized by the wave
vector $\textbf{k}_{\textbf{l}}$ of this wave. It will be
discussed  in Sec. 8 that this wave can be considered as a
particular kind of longitudinal sound. Therefore, the condensate
of $N$ ``zero-phonons'' is present in the ground state. Thus, the
resonance wave forms a crystal and supports the condensate of
atoms. In a fluid, the condensate is also formed by a wave, but
with $\textbf{k}=\textbf{k}_{1}$ (see I and  Ref.~\onlinecite{zero-cond}).

Wave function (\ref{1-6}) describes a simple rectangular lattice.
The more complicated lattices can be constructed, by setting a
relevant bare WF $\Psi^{bare}$ instead of the product of sines
$\Psi^{bare}_{sc}$ (\ref{2-5}). The important point is whether
$\Psi^{bare}$ reflects the structure of a Wigner--Seitz cell,
i.e., whether it equals zero on its surface. This is true for the
sc lattice, but is not necessarily for other lattices. If
$\Psi^{bare}\neq 0$ on the cell surface (we denote
$\Psi^{bare}_{cell}\neq 0$), then it is easy to guess the form of
$\Psi^{bare}$ for the bcc and fcc lattices:
\begin{eqnarray}
\Psi^{bare}_{bcc} &=& \Psi^{bare}_{sc}(N/2) \prod\limits_{j=1+N/2}^{N}\left \{\sin{(k_{l_{x}}(x_{j}-a_{l}/2))}
\right. \nonumber \\ &\times &\left. \sin{(k_{l_{y}}(y_{j}-a_{l}/2))}
 \sin{(k_{l_{z}}(z_{j}-a_{l}/2))}\right \} \nonumber \\ &+& \mbox{permutations}.
  \label{c-8}    \end{eqnarray}
  \begin{eqnarray}
 \Psi^{bare}_{fcc}&=& \Psi^{bare}_{sc}(N/4)\Psi^{xy}_{z}(N/4)\Psi^{zx}_{y}(N/2)\Psi^{yz}_{x}(3N/4) 
 \nonumber \\ &+& \mbox{permutations}.
  \label{c-9}    \end{eqnarray}
  \begin{eqnarray}
\Psi^{xy}_{z}(N_{0})&=& \prod\limits_{j=N_{0}+1}^{N_{0}+N/4}\left \{\sin{(\sqrt{2}k_{l_{x}}\acute{x}_{j})}
\sin{(\sqrt{2}k_{l_{y}}\acute{y}_{j})} \right. \nonumber \\ 
&\times & \left. \sin{(k_{l_{z}}[z_{j}+a_{l}/2])}\right \},
  \label{c-10}    \end{eqnarray}
  \begin{equation}
\acute{x}_{j}=(x_{j}-a_{l}/2)\cos{(\pi/4)}+y_{j}\sin{(\pi/4)}, 
  \label{c-11}    \end{equation}
  \begin{equation}
 \acute{y}_{j}=-(x_{j}-a_{l}/2)\sin{(\pi/4)}+y_{j}\cos{(\pi/4)}.
  \label{c-12}    \end{equation}
Here, the permutations symmetrize the WF to the Bose form, $a_{l}$
is the period of a lattice, and four functions in (\ref{c-9}) on
the right-hand side depend on the coordinates of atoms with the
numbers $1, \ldots, N/4$; $N/4+1, \ldots, N/2$; $N/2+1, \ldots,
3N/4$ and $3N/4+1, \ldots, N$. These WFs imply that, for the bcc
lattice,
$\textbf{k}_{c}=\pi(\textbf{i}_{x}+\textbf{i}_{y}+\textbf{i}_{z})/a_{l},$
and the condensate $\textbf{k}$ are the same as that for the sc
lattice. For the fcc lattice, there are four composite
condensates: with
$\textbf{k}_{c}=\pi(\textbf{i}_{x}+\textbf{i}_{y}+\textbf{i}_{z})/a_{l}$
and
$\textbf{k}^{\prime}_{c}=\pi(\sqrt{2}\textbf{i}_{\acute{x}}+\sqrt{2}\textbf{i}_{\acute{y}}+\textbf{i}_{z})/a_{l}$
with permutations of $x, y, z$.

 At  $\Psi^{bare}_{cell}= 0,$ $\textbf{k}_{c}$ are different.
For the sc lattice, the basis vectors of the reciprocal lattice are $\textbf{b}_{1}=2\pi\textbf{i}_{x}/a_{l}$,
$\textbf{b}_{2}=2\pi\textbf{i}_{y}/a_{l}$, and $\textbf{b}_{3}=2\pi\textbf{i}_{z}/a_{l},$ and the relation $\textbf{k}_{c}=\textbf{g}^{3D}_{0}/2
\equiv (\textbf{b}_{1}+\textbf{b}_{2}+\textbf{b}_{3})/2$ holds. For the bcc lattice, we have $\textbf{b}_{1}=2\pi(\textbf{i}_{y}+\textbf{i}_{z})/a_{l}$,
$\textbf{b}_{2}=2\pi(\textbf{i}_{x}+\textbf{i}_{z})/a_{l}$, and $\textbf{b}_{3}=2\pi(\textbf{i}_{x}+\textbf{i}_{y})/a_{l}$,
and the relation $\textbf{k}_{c}=\textbf{g}^{3D}_{0}/2=2\pi(\textbf{i}_{x}+\textbf{i}_{y}+\textbf{i}_{z})/a_{l}$.

We note that the solution for a one-particle WF with spherical
orbit was studied in Ref.~\onlinecite{krai2011}. There, it was assumed that
the atoms of the condensate have  $k\sim 2.1\,\mbox{\AA}^{-1}$.
This is close to our results.

It is of importance to confirm the existence of the condensate in
experiments. A neutron or a photon are scattered with the creation
of a quasiparticle or elastically. In the second case, the wave
vector of the scattering atom either is not changed or changes
only the direction (to the opposite one), and the state of the
crystal is invariable (except for the recoil). If the condensate
is present, then the scattering with the momentum transference
$2\textbf{k}_{c}$
 (or with a change in one-two components of the vector $2\textbf{k}_{c}$) gives the intense peak. At $\Psi^{bare}_{cell}= 0$ and at
 $\Psi^{bare}_{cell}\neq 0,$ these changes in the momentum are equal to the vector $\textbf{g}$ of the reciprocal lattice, and
the scattering corresponds to the Bragg--Wolf (BW) peaks. But if $\Psi^{bare}_{cell}= 0,$ then the number of condensate peaks is less than that of BW peaks.
At $\Psi^{bare}_{cell} \neq 0,$ the intensity of all BW peaks must increase at small $T$ (in the presence of a condensate)
proportionally to the amount of the condensate.
For bare WFs (\ref{2-5}), (\ref{c-8})--(\ref{c-10}), the BW peaks (noncondensate ones) can be, apparently, interpreted
as a result of the scattering by zero-phonons with $\textbf{k}=\textbf{k}_{c}$.

For the bcc, fcc, and hcp lattices, the Wigner--Seitz cell is complicated. Moreover, at $\Psi^{bare}_{cell}= 0,$ its creation requires a wave with complicated structure.
But, at $\Psi^{bare}_{cell}\neq 0,$ the lattice can be formed from waves with a simpler structure of the type (\ref{c-8})-(\ref{c-11}).
Since the Nature uses simple structures, we suppose that $\Psi^{bare}_{cell}\neq 0$.

We note that $\textbf{k}=\textbf{k}_{c}+\delta\textbf{k}$ for the higher condensates, and the components $\delta\textbf{k}$ are multiple to the
components $2\textbf{k}_{c}.$
Therefore, all condensates must be revealed in experiments as a part  of the ``base'' condensate with $\textbf{k}=\textbf{k}_{c}$.

\section{Comparison with the traditional approach}
Let us compare the wave solution (\ref{1-6}) with the traditional
one (\ref{1-1}). In the last case, it is assumed that the atoms
carry on small oscillations near  lattice points.

The traditional approach involves several assumptions. 1) The
lattice is set ``by hands'', it is not obtained from the
Schr\"{o}dinger equation. 2) It is assumed that the probability
density maximum in a cell is located at the lattice point, near which the
atom carries on random oscillations (atom is fixed likewise on
rubber string, which role is played by the function $\varphi
(\textbf{r}-\textbf{R})$). Here, one more courageous assumption is
hidden: that the mechanism of appearance of a lattice does not
influence the motion of atoms in it.
3) It is assumed also that the boundaries have no effect on the
solutions. Therefore, the realistic BCs close to
zero ones are replaced by the cyclic conditions.

Solution (\ref{1-6}) has no above-mentioned drawbacks. It is found
with the use of natural zero BCs. The lattice and
the probability distribution in a cell do not postulated, but they
follow from the solution of the Schr\"{o}dinger equation. It is
seen from the probability distribution that the motion of atoms is
strongly affected by the mechanism (wave one) of formation of a
crystal. In addition, the observable quantities, in particular
$E_{0}$ and the energy of phonons, depend significantly on the
BCs. This is caused by the fact that the walls
change the Fourier expansion of the two-particle potential (see I:
formulas (15)-(17) and Sec. 5) and affect the phonon frequency
through $\Psi_{0}.$ Visually, it is because the excitations of 
the quantum system are a standing waves  
rather then a particles; a  wave keeps the memory  about the wall and 
its length is modulated by the  wall.
Such influence exists only for
natural long-range potentials.
In the above equations for $E_{0}$ and $E(k),$ this influence is
manifested in the factor $1/8$ at the potential $\nu_{3}(k)$ and
in the summation over the wave vectors multiple to $\pi/L$, rather
than $2\pi/L,$ in a  corrections. To the effect of boundaries, we
can refer also the corrections $Q_{j}$ (changing the phonon
frequency $E(k)/\hbar$), which arise from the product of sines in
(\ref{1-6}) and are induced by the interaction of a phonon with $N
$ zero-phonons of the ground state.

Relation (\ref{1-6}) yields easily the traditional solution
(\ref{1-1}). Let each atom be localized near a lattice point, one atom per
point. The points $\textbf{R}_{j}$ correspond to the extrema of a
sine. By expanding the sines in a Fourier series near points, the
terms linear in $\textbf{r}_{j}-\textbf{R}_{j}$ are absent, and
the small quadratic terms can be taken up in the exponent. Thus,
for the sc lattice, we have
\begin{eqnarray}
 &&\prod\limits_{j=1}^{N}\{\sin{(k_{l_{x}}x_{j})}\sin{(k_{l_{y}}y_{j})}
 \sin{(k_{l_{z}}z_{j})}\} \nonumber \\ 
 &\rightarrow &\pm\prod\limits_{j=1}^{N}e^{-\alpha^{2}(\textbf{r}_{j}-\textbf{R}_{j})^{2}/2},
  \label{5-1}    \end{eqnarray}
where  $\alpha=k_{l_{x}}=\pi/\bar{R}$
(the close estimate $\alpha \simeq 1-2\,\mbox{\AA}^{-1}$ was obtained in Ref.~\onlinecite{rev1} from the other reasoning).
By setting $S_{w}^{(\textbf{l})}=0$ in (\ref{1-6}), we reduce (\ref{1-6}) to (\ref{1-1}).

Thus, the traditional solution (\ref{1-1}) is a simplification of
the wave solution (\ref{1-6}). It is significant that, in this
case, the wave character of the solution is lost; but since the
counting-off is made from the equilibrium positions of atoms, it
is sufficiently simple to calculate 
\cite{rev1,saunders1962,mullin1964,nosanow1964,nosanow1966,nosanow1962,nosanow1967,glyde1976}
(with several fitting parameters) $E_{0}$ and $E(k)$.
In this respect, approach (\ref{1-1}) has certain advantages,
since the wave approach (\ref{1-6}) yields a chain of complicated
equations with many large corrections for $E_{0}$ and $E(k).$
These equations cannot be solved exactly.

The frequency of phonons $E(k)/\hbar$ is usually calculated in the
harmonic approximation, where a phonon is a wave arising at a
small deviation of atoms from equilibrium positions. However, it
was noted above that the phonon frequency is affected by
zero-phonons. In the language of oscillating atoms, this means
that the atoms in the ground state are not in rest at lattice points (as is
considered in the harmonic approximation), but they oscillate
intensively due to the motion in the field of zero-phonons. In
this case, the potential energy is minimum, probably, at lattice points.
 But it would be wrong to identify the deviations from
``equilibrium positions'' with the coordinates of real atoms,
since, in this case, the intense zero oscillations would be lost.
Therefore, the harmonic approximation is applicable only for the
description of long-wave oscillations, when a crystal can be
considered as a continuum. This approximation cannot be used for
short-wave phonons, and the corresponding approaches in
solid-state physics should be reconsidered, in our opinion. The
agreement with experiment of phonon dispersion curves calculated
in the harmonic approximation in a number of works  seems to be
accident or is due to the choice of parameters.

WF (\ref{1-1}) is not a solution of the Schr\"{o}dinger equation.
The attempt to determine $\varphi (\textbf{r}-\textbf{R})$ in
(\ref{1-1}) from the Schr\"{o}dinger equation leads \cite{rev1} to
the solution $\varphi (\textbf{r}-\textbf{R})\sim
e^{i\textbf{k}\textbf{r}}$, which is nonlocalized and, therefore,
unphysical.  It was assumed \cite{rev1} that this liquid-like
solution arises due to the truncation of the cluster expansion.
However, instead of $e^{i\textbf{k}\textbf{r}},$ we can take a
linear combination of exponents $\varphi
(\textbf{r}-\textbf{R})\sim \sin{(k_{l_{x}}x)}\sin{(k_{l_{y}}y)}
 \sin{(k_{l_{z}}z)}$. The we arrive at the wave solution (\ref{1-6}), which is quite physical
and sets a lattice with localized distribution of atoms.

Recently, N. Prokof'ev \cite{rev3} considered a wave solution of the form
\begin{equation}
 \Psi=\prod\limits_{i=1}^{N}\left (\frac{1}{\sqrt{N_{L}}} \sum\limits_{j=1}^{N_{L}}\varphi(\textbf{R}_{j}-\textbf{r}_{i})\right ),
  \label{5-4}    \end{equation}
which contains a condensate in the state
$\frac{1}{\sqrt{N_{L}}}\sum\limits_{j=1}^{N_{L}}\varphi(\textbf{R}_{j}-\textbf{r})$.
However, this solution was recognized unphysical \cite{rev3},
because the number $N_{L}$ of the lattice points can differ from
the number $N$ of atoms. Hence, the solution for a crystal is only
one of the huge number of solutions, and the probability of its
realization is too small. We note that $N_{L}$ in (\ref{1-6}) can
also be different from $N$. However, it is easy to see (see the
following section) that we have no problems in this case.

Thus, solution (\ref{1-6}) can be obtained long ago.

In the last time, the following new approaches to the description
of quantum crystals are developed: Path Integral Monte-Carlo
method \cite{ceperley2006}, variational Shadow WFs method
\cite{swf1998} using the bare WF (\ref{1-1}) and many fitting
parameters, and Shadow Path Integral Ground State projector method
\cite{spigs2006}. The last method used the Bijl-Jastrow function,
and the lattice arises due to the spontaneous breaking of
symmetry.

\section{Ground state --- liquid or crystal?}
It is accepted that most substances in the ground state are
crystals. Apparently, the ground state corresponds always to a
liquid.

The question about the structure of the ground state can be
answered, by seeking a minimum of the energy. Consider a sc
crystal. Its ground-state energy is given by formula (\ref{2-31}).
In (\ref{2-31}), we fix the number of atoms $N$ (i.e., the value
of $\bar{R}$) and change the number of points $N_{L}$ (i.e., the
value of $k_{c}$). In the zero approximation (\ref{2-27}) for
$a_{2}(k),$ the curve $E_{0}(k_{c})$ is a parabola (see Fig. 4),
where the point $k_{c}\approx 1.41\,\mbox{\AA}^{-1}$ corresponds
to $N_{L}=N$. If we descend somewhat downward along the curve, we
obtain the states with $N_{L}<N$, where some cells have two atoms.
In such cells, the distance between two atoms must be small,
$\simeq \mbox{\AA}.$ Since the atoms have the almost hard core
with a radius of $1.2-1.3\,\mbox{\AA},$ they strongly repel each
other at distances of $\lsim 2.5\,\mbox{\AA}.$ Therefore, such
solution is unstable: one of the atoms leaves the cell and will
walk in the crystal, until it approaches the crystal surface.
There, the atom evaporates or becomes fixed. As a result, the
system transits in the state with $N_{L}=N$. If we lift upward
along the curve, we obtain the states with vacancies ($N_{L}>N$).
Such states have higher energies. Hence, the stable state of the
crystal with minimum energy has $N_{L}=N$, and each cell contains
exactly one atom. This reasoning answers the objection in
Ref.~\onlinecite{rev3}.
\begin{figure}[h]
\centerline{\includegraphics[width=85mm]{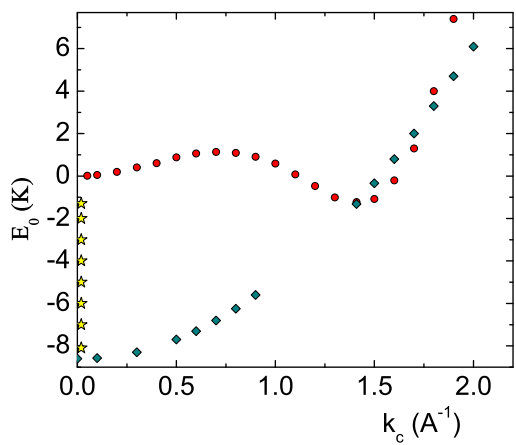}}
\caption{Function $E_{0}(k_{c})$ for a sc lattice of $He^4$ atoms
at the fixed $\bar{R}=3.85\,\mbox{\AA}$ (rhombs) and the variable
$\bar{R}$ under the condition $N_{L}=N$ (circles). Stars show the
underliquid at various temperatures. We use the potential
(\ref{2-28}) with $a= 2\,\mbox{\AA}$, $b= 4\,\mbox{\AA}$,
$U_{bd}=-9\,$K and $U_{b} =81.2\,$K. \label{fig4}}
\end{figure}

The curve of rhombs indicates that
$E_{0}(k_{c}=1.6\,\mbox{\AA}^{-1})\approx E_{0}(k_{c}=1.41\,\mbox{\AA}^{-1})+1.1\,$K. At
 $k_{c}\approx 1.6\,\mbox{\AA}^{-1},$ the system has $0.46$ vacancy per atom. Hence, the vacancion energy $E_{v}\approx 2.4\,$K, which is comparable, by the order of
magnitude, with experimental value \cite{goodkind1990} $E_{v}\sim 10\,$K.

The dependence $E_{0}(N)$ (circles in Fig. 4)
is quite reasonable: the most favorable state of the lattice corresponds to the minimum
($k_{c}\approx 1.41\,\mbox{\AA}^{-1}$, $\bar{R}\approx 3.85\,\mbox{\AA}$).
By passing to a rarefied system ($\bar{R}\rightarrow \infty, k_{c}\rightarrow 0$),
we have $E_{0}\rightarrow 0$.

The minimum of the parabola of rhombs corresponds to the least
possible $k_{c}=k_{1}=\sqrt{3}\pi /L$, at which WF (\ref{1-6})
transits in $\Psi_{0}$ of a fluid, and $E_{0}$ (\ref{2-31}) is
determined only by the term $\tilde{E}^{b}_0$ for fluids. Hence,
the ground state of a fluid ($E_{0}\approx -8.6\,$K) lies below
the ground state of the crystal ($E_{0}\approx -1.27\,$K). The
states on the parabola near the minimum (e.g., with $k_{c}\sim
0.1\,\mbox{\AA}^{-1}$) correspond to a fluid partitioned into
domains with many atoms. Such states must be unstable.

We have studied Eq. (\ref{2-31}) obtained in a rather crude
approximation. Its solution can be changed with regard for
corrections and the exact potential (with a higher barrier). It
can turn out that the ground state includes the admixture of
vacancies \cite{andreev1969}, though this is improbable.

The main question is as follows: \textit{Is $E_{0}$ of a crystal
greater than $E_{0}$ of a liquid or less, for the exact solution?}
The answer is hinted by the node theorem \cite{gilbert}: ``if the
eigenfunctions of a self-adjoint second-order differential
equation are arranged in some region $G$ under any homogeneous
BCs in the increasing order of the appropriate
eigenvalues, then the nodal manifolds of the $n$-th eigenfunction
$\Psi_{n}$ divide the region $G$ into at most $n$ subregions for
any number of independent variables''. The theorem was proved for
WFs of the general form $\Psi(\textbf{r}_1,\ldots ,\textbf{r}_N)$.
We deal with the WFs of $N$ identical Bose particles, and the
conditions of the theorem are satisfied. We know that WF
(\ref{1-6}) with $\textbf{k}_{\textbf{l}}\rightarrow
\textbf{k}_{1}$ corresponds to the ground state of a liquid and
has no nodes. By the theorem, namely this function describes the
ground state of a system of Bose particles in a box. Hence, all
remaining states, including the ground state of the crystal, have
nodes and correspond to higher energies. But it can be so that the
equations have no solutions for $\Psi_{0}$ corresponding to a
liquid. Does $\Psi_{0}$ of the crystal without nodes exist? It is
clear that, at the wave structure of (\ref{1-6}), the WF of the
crystal has necessarily nodes. Can a nonwave structure be
realized? The WF must turn to zero at the surface of the crystal
and be symmetric under permutations of atoms. It is easy to see
that the nonwave solution (\ref{1-1}) is impossible without nodes:
it does not feel the boundaries and can become zero only at
infinity. We cannot strictly prove this assertion, but we sure
that only the wave solution satisfies the zero BCs. This implies that 1) the ground-state energy of the
crystal is always higher than that of a liquid composed of the
same atoms; 2) the solution for $\Psi_{0}$ describing a liquid
exists necessarily for Bose atoms of all sorts (if the solution is
absent at some density $\rho$ of a liquid, it is possible to enter
a region, where a solution exists, by varying $\rho$).

However, the majority of substances in the Nature become crystals
at low temperatures. The transition in the crystal state allows
the system to decrease the energy by jump and is favorable. But it
is seen in Fig. 4 that the principal minimum corresponds to a
liquid. Apparently, by ``having fallen'' in the crystal state, the
system cannot already pass to a deeper minimum corresponding to a
liquid. To make this, the system must overcome a band of unstable
state (or the energy barrier that can appear at the exact
solution). Therefore, the majority of substances at low
temperatures are crystals.

The state of a liquid in the deep minimum can be called the
underliquid (UL, stars in Fig. 4). Most probably, the majority of
ULs has the superfluid phase. The temperature (determined by
quasiparticles) of UL can be equal to the temperature of a
crystal, but the total energy  is less than that of the crystal.
$He^4$ atoms have a large amplitude of zero oscillations.
Therefore, the lattice is locally unstable, apparently, and
transits in the UL state. In other words, He II is the single
example of UL among inert elements. Possibly, other substances can
be also transferred in the UL state (see Sec. 8). The amorphous
bodies with microstructure of a liquid are not related,
apparently, to UL, since the amorphous state is caused by a strong
anisotropy of molecules, whereas we have considered the systems of
spherical molecules.

We arrive at a significant conclusion that a finite system of Bose
particles of any sort (He, Ar, Ne, etc.) in the lowest state is a
liquid, rather than a crystal, as is commonly accepted. Sometimes,
the third law of thermodynamics and the entropy-based arguments in
favor of a crystal are discussed. However, the entropies of a
crystal and a liquid in the ground state are identical and are
equal to zero: $S=k_{B}\ln{N_{s}}=k_{B}\ln{1}=0$. In other words, the
degrees of order of a fluid  and a crystal are identical in the
ground state, though a crystal seems visually to be more ordered.

\section{Nature of the supersolid phase}
Almost all researchers arrived at the consensus \cite{rev3,spigs2006,superglass,grain,rota2011,ceperley2006,chan2008}
of that the supersolid phase and NCIM \cite{kim-chan1,kim-chan2} are related to defects of the lattice.
However, the attempts to identify a carrier of NCIM with a specific defect met difficulties, which is not surprising.
How can a crystal contain so many defects (or so extended defects) at ultralow temperatures $T\simeq 0.02\,$K
that they connect $20\%$ of atoms of the lattice,
by ensuring the experimental value $\rho_{s}\simeq 0.2\rho$?
In our opinion, it is improbable. We note that the defects are an analog of quasiparticles; for comparison:
the amount of quasiparticles in He II at $T=0.02\,$K is so small that they provide $\rho_{n} \sim 10^{-7}\rho$.
Therefore, we suppose that the carriers of the effect are atoms of the ideal lattice.

Let us consider various possibilities in detail. We start from
vacancions. Since any crystal is in the gravity field, the
vacancions undergo the action of the force directed upward. If the
gas of vacancions is superfluid, then the vacancions should float up
rapidly and evaporate from the surface. Under torsional
oscillations, they must be transported to an internal surface of a
crystal.
Since this reasoning is
valid for any massive defects, NCIM is not related to vacancions.

The models of dislocation glass \cite{disloc-glass},  superglass \cite{superglass},
and grain boundaries \cite{grain}
assume the existence of the condensate of atoms with $k=0$, but it was not observed \cite{diallo2011}: $n_{0}\lsim 0.003$. In the model of
screw dislocation network \cite{disloc-net}, the superfluid component consists of atoms of the nuclei of dislocations, whose amount is obviously much less than
10\%. Therefore, this model does not explain the observed large values of $\rho_{s}\simeq (0.1-0.2)\rho$. The models of
dislocation network met the analogous difficulty \cite{shevch1987,disloc-glass}.
As for the mechanism of grain boundaries \cite{grain}, it does not agree with the fact that NCIM was observed in a monocrystal \cite{chan2007} (where any grains are absent).

As carriers can be atoms of the lattice.  It was proposed already
\cite{krai2011}, but no mechanism of correlations (of the
condensate) was indicated. We propose the following scenario. For
the sc lattice, $\Psi_{0}=0$ on the boundary of a Wigner--Seitz
cell. Therefore, the atom cannot pass from a cell to another one.
However, for the bcc, fcc, and hcp lattices, $\Psi_{0} \neq 0$
most probably on the boundary of a cell (see Sec. 4), and the
tunneling of an atom from cell to cell is possible. Consider a bcc
crystal. According to Sec. 2, the atom is located near the
orbit, whose size is twice less than that of a cell of the
crystal. The orbits of the atom at the center of a cubic cell and
the atom of one of eight vertices of a cube touch each other.
Therefore, the WFs of such atoms are strongly overlapped. If the
condensate includes more than a quarter of atoms, then each atom
is contiguous to two or more atoms of the condensate. In this
case, many neighboring atoms of the condensate can be joined by
lines. Along such closed lines, the atoms can flow by means of the
simultaneous tunneling (according to the structure of WF
(\ref{c-8}), one atom cannot flow through a crystal, since the
zeros of a sine create the impermeable planes in a certain
distance). Due to the indistinguishability of atoms, the atoms
belonging to the condensate and the overcondensate change places.
Therefore, all atoms participate in the flow. Since the atoms of
the condensate are correlated, they can move as the whole, by
representing the flowing component of a crystal. In order that
this component be superfluid, its excitations must satisfy the
Landau criterion. The dispersion curve of such excitations was
apparently observed \cite{sf-sp1,sf-sp2}, and it satisfies the
Landau criterion. We assume that the superfluidity is possible
only in the case where the concentration $n_{c}$ of the condensate
is higher than the threshold one ($n^{cr}_{c}= n/4$ for the bcc
lattice). Sec. 4 indicates that such $n_{c}$ is possible.

Let us turn to the optic-like dispersion curves \cite{sf-sp1,sf-sp2}. They are very close, and we can have no doubts that they represent the same mode.
In Ref.~\onlinecite{sf-sp1}, this mode was considered vacancional. But the experiment \cite{sf-sp2} with a polycrystal showed that it
disappears at $0.2\,\mbox{K}< T < 0.6\,$K.  Though, the ordinary vacancional mode must be present at all $T$;
it shifts only, as the density varies \cite{goodkind1990}.
In Ref.~\onlinecite{sf-sp2}, the mode is referred to the superfluid component, and we agree 
with it. However, it is the mode related to a crystal (rather than to a liquid \cite{sf-sp2}),
because the minimum is just at the Brillouin zone boundary. According to our approach, this superfluid mode
must appear at the cooling
of a crystal down to the temperature, where the tunneling of the condensate starts.
It is $T_{NCIM},$ the temperature, where the effect of NCIM arises. This  is in agreement with data  \cite{sf-sp2}.
In Ref.~\onlinecite{sf-sp1}, the mode was observed also at $T = 0.6\,$K, which is more than
the known $T_{NCIM}$. But, in that case, a monocrystal was used. For it, $T_{NCIM}$ is unknown and can exceed $0.6\,$K.
We mention one more optic-like mode \cite{markovich2002} observed for the bcc lattice of $He^{4}$.
It should be noted that the data in Refs.~\onlinecite{sf-sp1,sf-sp2,markovich2002} say nothing about the nature of the superfluid component, which can be vacancional.

We note that since the flow can occur only through  lattice points, and some condensate chains can break,
the atoms of the condensate must be  involved in torsional oscillations of a crystal only partially. So that NCIM is proportional to
 $qn_{c}$, where $q< 1$. The value of $q$ is decreased by various defects. Therefore, $q$ must strongly depend on
the conditions of an experiment (this is observed), and $\rho_{s}$ can be small at high $n_{c}$. The maximum of $q$ for the
crystals with a large ratio of area to volume
\cite{reppy2008} can be explained by the disappearance of large defects hampering the flow of the condensate.
It is clear that the effect is maximum for the ideal crystal without admixtures and defects.

We note that the destruction temperature for the condensate must be significantly higher than
$T_{NCIM}$, since the tunneling flow requires $n_{c}\geq n^{cr}_{c}$.

The proposed tunneling mechanism agrees with the absence of the
superfluidity of crystals at a pressure gradient
\cite{beamish2006,sasaki2007}. According to
Refs.~\onlinecite{beamish2006,sasaki2007}, the superfluidity should be
accompanied by a deformation of the lattice (if defects are
absent). But, at the tunneling mechanism, the atoms can move in
the lattice without any change of its shape. Namely the torsional
oscillations \cite{kim-chan1,kim-chan2} present the conditions for
such motion. At the tunneling mechanism, the crystal cannot also
shift as a whole, since, to realize it, the surface atoms must
tunnel outside of the crystal, where there are no 
lattice points. For the same reason, one more possible mechanism
\cite{sasaki2007} such as the leakage of atoms of a fluid through
a crystal is forbidden. All this corresponds to the conclusion
\cite{beamish2006} that the mechanisms of flows for a crystal and
a fluid are quite different.

A decrease of NCIM after the annealing \cite{rittner2006} can be related not to a decrease in the number of defects,
but to the appearance of lengthy defects. It is well known that a bullet piercing glass makes only a small hole, whereas a stone
creates additionally a network of long cracks.
In other words, the slow processes are accompanied by extended deformations of a crystal. Under the annealing, a crystal is firstly strongly heated
and then is slowly cooled.
In this case, the majority of defects disappear, but the remaining dislocations are ordered into a three-dimensional network \cite{kittel-vved},
which hampers the flow of the condensate to a higher degree than many disordered dislocations.

The effect of NCIM was mainly observed in polycrystals. The tunneling of atoms in them
is possible, since there are no slits between microcrystals.

The experimentally found \cite{ftint2008} law $\triangle p \sim T^2$  (it is equivalent to $C_{v}\sim T$) can be because
the condensate density is close to the threshold value: $n_{c}\simeq n^{cr}_{c}$.
In this case, the dimension of the network of lines, by which the condensate flows, is close to 1.

The jump of the rigidity of a crystal \cite{day2007,kim2011} at
$T\approx T_{NCIM}$ means that the same factor affects the
rigidity and NCIM. We assume that this factor is the appearance of
the superfluid component. As a result, a part of energy is
transferred to modes of the superfluid component, and there occurs
a redistribution of oscillatory modes of the system. This causes
the jump of the rigidity. An increase of NCIM \cite{chan2008-2}
due to $He^3$ atoms can be related to the fact that $He^3$ atoms
join dislocations and, under the action of the inertial force at
torsional oscillations, turn a dislocation or move it to the
surface of a crystal.

It is necessary to understand why the heat capacity peak is less for purer crystals, and the temperature $T_{C_{v}}$
of the peak is less than $T_{NCIM}$ and is independent of the
concentration of the $He^3$ admixture \cite{chan2009} (though $T_{NCIM}$ depends strongly on it).
The first property is similar to the annealing effect. We assume that the reason is the same: purer (on the average)
crystals contain more lengthy defects.
Consider now the second and third properties. The ``excess'' of $C_{v}$ is related to the superfluid subsystem
and increases approximately linearly \cite{chan2009} at low $T$.
As $T$ increases, a part of chains, on which the superfluid flow is realized,
is broken due to the approach of $n_{c}$ to the threshold and the freezing-out of defects.
Let a half of chains be broken. If many alternative chains remain, then NCIM is almost not changed,
but the number of modes of the superfluid subsystem (i.e., $C_{v}$ also) decreases twice. NCIM decreases sharply only if
the number of chains becomes so little that the atoms lose the ability to flow. Apparently, as $T$ decreases, the superfluidity arises gradually:
first, a small number of atoms of the condensate can flow.
In this case, NCIM and the peak of $C_{v}$ are not related to the phase transition,  $T_{NCIM}> T_{C_{v}}$, and the peak of $C_{v}$
is independent of the $He^3$ admixture and is caused by the dependence of the number of phonons in the condensate
network of atoms on the properties of this network.

As is seen, the proposed model can explain the supersolid
phenomenon, but remains many questions unsolved. It should be
clarified whether $\Psi_{0}$ becomes zero on the surface of a
Wigner--Seitz cell for the bcc and hcp lattices. It is also
necessary to know which and how many defects are present in a
crystal under various conditions, and how each sort of defects
influences the condensate and its fluidity. The key moment for the
model is the experimental discovery of a
composite condensate with  $\textbf{k}=\textbf{k}_{c},
3\textbf{k}_{c}, 5\textbf{k}_{c}$.

 \section{Discussions}
It is seen from formula (\ref{1-6}) that a crystal is formed by a standing wave in the probability field.
A fluid has an analogous wave, but with $\textbf{k}=\textbf{k}_{1}=(\pi/L_{x}, \pi/L_{y},
\pi/L_{z})$. In I, it is shown that the wave in a fluid is similar to $N$ standing sound waves
with $\textbf{k}=\textbf{k}_{1}$. The same arguments are true
also for a crystal. Therefore, we can assert that a crystal in the ground state has $N$ identical
standing longitudinal acoustic phonons with $\textbf{k}=\textbf{k}_{c}$.
They are particular resonance zero-phonons. By the structure of corrections, they differ from ordinary phonons described by WF
(\ref{3-1})--(\ref{3-4}). It is significant that these standing zero-phonons create a crystal lattice.
In other words, the \textit{periodicity of a crystal is caused by the periodicity of a sound wave}.

The number of resonance phonons is equal to the number of atoms. At such huge occupation number, these phonons can be considered as a single classical sound wave.
But if a crystal is a wave, we can try to control its state with the help of sound and
electromagnetic waves.
In particular, it would be possible to create or to destroy crystals.

Well-known are the legends concerning N. Tesla \cite{tezla}, who induced vibrations
of building's walls with the help of a small mechanical oscillator with the resonance frequency
in the ultrasound region. Apparently, N. Tesla excited the eigenmodes with $\textbf{k}$
(\ref{2-2}). The particular resonance frequency for crystals is the frequency of zero-phonons,
which is equal to the difference of $E_{0}/\hbar$ of a crystal and $E_{0}/\hbar$ of a fluid. The wave with such a frequency forms a crystal itself.

In Sec. 6, the state of underliquid is predicted. Possibly, this state can be obtained by means of the wave action
on a crystal or a rapidly cooled fluid (so that the fluid will avoid the state of crystal and become UL). Through a liquid, it is necessary to
transmit monochromatic sound or electromagnetic waves
with a wavelength comparable (but not equal to) with the period of a lattice. It is possible to use an x-ray or gamma-laser
with $0.5\,\mbox{\AA} \lsim\lambda \lsim 10\,\mbox{\AA}.$ But such lasers have not been created till now, to  our knowledge.
A crystal can be destroyed,
possibly, by a wave with the frequency of a zero-phonon.
All this remains else at the level of speculations, but it  is interesting to study these quastions.

The crystallization of a liquid can be related to resonance phenomena in the system of
phonons. The centers of crystallization are usually considered in the language of interacting atoms.
But, according to the wave solution, these centers are, most likely, growing wave packets
with $\textbf{k}=\textbf{k}_{c}$.

We assume that the wave principle is the general one for the
formation of crystals. As an important task, we indicate the
search for the solutions for the bcc, fcc, and hcp lattices, which
are most spread in the Nature.

\section{Conclusions}
Our most important conclusion concerns the wave nature of Bose
crystals. The wave properties yield the condensate of atoms and
the possible superfluidity of a crystal. If the crystals in the
Nature are created by standing waves, it is quite beautiful.

The complexity of properties of quantum crystals is related to the fact that they have five subsystems:
atoms of the lattice, atoms of the condensate, quasiparticles in both these systems, and various defects.
This makes it difficult clarifying the nature of the supersolid phase.
Though He II has only two subsystems, the superfluid subsystem and that of quasiparticles, the nature of its superfluidity was understood in more than one decade.

The properties of quantum crystals are fine and arouse the feeling of admiration.
In addition, it seems clearly that Lady Science moves away from Truth sometimes in her walks.
But She goes not so quickly, as  the Universe expands, and can return always.

The present work is devoted to the memory of Petr Ivanovich Fomin.

\section{Appendix. Calculation of $C_{l}(j)$}
Formula (\ref{2-8}) arises from the expansion of
 $\tilde{f}(x)$ (\ref{2-f}) in a Fourier series as $\delta \rightarrow 0$. Instead of $\tilde{f}(x),$ we can take any function, which satisfies the conditions for
the Fourier expansion to exist and passes to $\cot{(k_{l_{x}}x)}$ as $\delta \rightarrow 0.$
Since the input function can be expanded in a Fourier series at $\delta \neq 0$,
the quantity $C_{l_{x}}(j_{x})$ can be determined by the formulas of Fourier analysis:
 \begin{eqnarray}
 & &  C_{l_{x}}(j_{x}) =  - C_{l_{x}}(-j_{x}) = \lim\limits_{\delta \rightarrow 0}
 \frac{1}{L_{x}}\int\limits_{0}^{L_{x}} dx\tilde{f}(x)e^{-i2\pi j_{x}x/L_{x}}\nonumber \\
 &=&\frac{1}{L_{x}}\int\limits_{0}^{L_{x}}
 dx\cot{(k_{l_{x}}x)}e^{-i2\pi j_{x}x/L_{x}} \nonumber \\ 
 &=& -2i\int\limits_{0}^{1/2}
 dx\sin{(2\pi j_{x}x)}\cot{(\pi l_{x} x)}.
     \label{A-1} \end{eqnarray}
Integral (\ref{A-1}) is defined, despite the discontinuities of a cotangent, and can be calculated in the sense of the principal
value of an improper integral \cite{whittaker1}.
By induction, relation (\ref{A-1}) yields
$C_{l_{x}}(j_{x}) =  -i $ at $j_{x}= l_{x},  2l_{x},  \ldots $, and
$C_{l_{x}} =  i $ at $j_{x}= -l_{x},  -2l_{x},  \ldots $.
However, the proof of the third part of (\ref{2-9}), namely the vanishing of $C_{l_{x}}$ at the rest $j_{x}$, is not a simple task.

We use the following trick. In I, it was shown that
\begin{equation}
 \cot{(\pi x/L)} = -i
 \sum\limits_{j\neq 0}(j/|j|)e^{i2\pi jx/L}.
     \label{A-2} \end{equation}
This equality should be understood in the same meaning as (\ref{2-8}), which was obtained in I from the
Fourier expansion of the smooth function $\cot{(k_{1x}x+\delta_{1x})}$ as $\delta_{1x} \rightarrow 0$.
Formula (\ref{A-2}) follows also from the sum of two geometric progressions:
\begin{eqnarray}
 \cot{(\pi x/L)} &=& \left \{-i
 \sum\limits_{j=1, 2, \ldots}e^{j(-\delta +i)2\pi x/L}\right. \nonumber \\
 &+& \left.\left. i\sum\limits_{j=-1, -2, \ldots}e^{j(\delta +i)2\pi x/L}\right \}\right |_{\delta\rightarrow 0},
     \label{A-3} \end{eqnarray}
where $\delta >0$, $x \in [0, L]$. Replacing $x \rightarrow l_{x}x$ in (\ref{A-2}) or (\ref{A-3}), we obtain (\ref{2-9}).


\renewcommand\refname{}





        \end{document}